\documentclass[aps,pra,reprint,superscriptaddress,amsmath,amssymb,amsfonts,showkeys]{revtex4-1}
\usepackage[pdftex]{graphicx}
\usepackage{dcolumn}
\usepackage{bm}
\usepackage{epsfig}
\usepackage{pifont} 
\usepackage{multirow}
\usepackage{subfigure}
\usepackage{float}
\usepackage[utf8]{inputenc} 
\usepackage{epstopdf}
\usepackage{MnSymbol}

\newcommand{\ba}{\begin{eqnarray}}
\newcommand{\ea}{\end{eqnarray}}
\newcommand{\be}{\begin{equation}}
\newcommand{\ee}{\end{equation}}
\newcommand{\bea}{\begin{eqnarray}}
\newcommand{\eea}{\end{eqnarray}}

\newcommand{\T}{\mathrm{T}}
\newcommand{\Tr}{\mathrm{Tr}}

\newcommand{\wnull}{W_{| 0 \rangle}}
\newcommand{\Qnull}{Q_{| 0 \rangle}}
\newcommand{\Pnull}{P_{| 0 \rangle}}

\newcommand{\spinvacuumpure}{| J\!J \rangle }

\newcommand{\Y}{\mathrm{Y}}

\newcommand{\abs}[1]{\ensuremath{\vert #1 \vert}}
\newcommand{\nproj}{N_r}

\DeclareMathOperator{\tr}{tr}

\def\i {\mathfrak{i}}

\def\N {\mathbb{N}}

\usepackage{bbm}

\usepackage{amsthm}

\newtheorem{result}{Result}

\usepackage[bookmarks=false,pdfstartview={FitH}]{hyperref}

\allowdisplaybreaks

\begin{document}

\title{Continuous phase-space representations for finite-dimensional quantum states\\ and their tomography}

\author{Bálint Koczor}
\email{balint.koczor@materials.ox.ac.uk}
\affiliation{Technical University of Munich, Department of Chemistry, Lichtenbergstrasse 4, 85747 Garching, Germany}
\affiliation{Munich Center for Quantum Science and Technology (MCQST), Schellingstrasse~4, 80799 München, Germany}
\affiliation{University of Oxford, Department of Materials, Parks Road, Oxford OX1 3PH, United Kingdom}
\author{Robert Zeier}
\email{r.zeier@fz-juelich.de}
\affiliation{Technical University of Munich, Department of Chemistry, Lichtenbergstrasse 4, 85747 Garching, Germany}
\affiliation{Forschungszentrum Jülich GmbH, Peter Grünberg Institute, Quantum Control (PGI-8), 54245 Jülich, Germany}
\author{Steffen J. Glaser}
\email{glaser@tum.de}
\affiliation{Technical University of Munich, Department of Chemistry, Lichtenbergstrasse 4, 85747 Garching, Germany}
\affiliation{Munich Center for Quantum Science and Technology (MCQST),
              Schellingstrasse~4, 80799 München, Germany}

\begin{abstract}
Continuous phase spaces have become a powerful tool for describing, analyzing, 
and tomographically reconstructing quantum states in quantum optics and beyond. 
A plethora of these phase-space techniques are known, however a thorough 
understanding of their relations was still lacking for finite-dimensional quantum states. 
We present a unified approach to continuous phase-space representations which 
highlights their relations and tomography. The infinite-dimensional case from quantum optics is then recovered 
in the large-spin limit.
\end{abstract}

\date{November 2, 2019}


\maketitle

\section{Introduction}
Phase spaces provide both theoretically and experimentally 
useful ways to visualize and analyze abstract states of 
infinite- and finite-dimensional quantum systems.
A plethora of phase-space representations are known 
\cite{SchleichBook,zachos2005,schroeck2013,Curtright-review}, including 
the Glauber P, Wigner, and Husimi Q function,
each of which has provided insights in 
quantum optics, quantum information theory, and beyond.
Phase spaces have also played an essential role
in characterizing the quantum nature of light
and became a natural language for quantum optics
due to the seminal work of Glauber \cite{Glauber1963,glauber2006nobel,cahill1969}, 
also clarifying
their interrelations in terms of Gaussian convolutions.
Beyond quantum optics, phase spaces are 
conceptually invaluable and provide a complete
description of quantum mechanics. They mirror
and naturally reduce to
classical phase spaces in the limit of
a vanishing Planck constant
\cite{Gro46,Moy49,1bayen1978,2bayen1978,berezin74,berezin75}.
Phase-space techniques and their associated quantizations \cite{Wey27,Weyl31,Weyl50}
have been widely applied in the context of
harmonic analysis and pseudo-differential operators
\cite{thewignertransform,deGosson2016,groechenig2001foundations,cohen1966generalized,Cohen95}.
In this work, we focus on  \emph{finite-dimensional}
quantum states, for which phase-space methods have been explored only to a lesser extent.

Recent advances in experimentally creating entangled
quantum states for spins or spin-like systems, such as
atomic ensembles \cite{mcconnell2015,haas2014},
Bose-Einstein condensates \cite{anderson1995,ho1998,ohmi1998,stenger1999,lin2011,treutlein2010,Schmied2011,hamley2012,strobel2014},
trapped ions \cite{leibfried2005,bohnet2016,monz2011},
and light polarization \cite{bouchard2016,klimov2017,chaturvedi2006}, 
have been in certain cases illustrated with
phase-space techniques and therefore 
call for a more profound understanding of these tools
with regard to
finite-dimensional quantum states.
To this end, we present a general approach to
continuous phase spaces for spins
which clarifies their interrelations 
by conveniently translating between them,
while emphasizing the connection 
to the infinite-dimensional case from quantum optics.
We do not consider discrete phase spaces such as the one proposed by
Wootters \cite{Wootters87}, see also
\cite{leonhardt1996,gibbons2004,ferrie2009} and references therein.

Phase-space representations have 
become crucial in the tomographic
reconstruction of infinite-dimensional quantum states \cite{Leonhardt97,SchleichBook}.
The optical homodyne tomography 
reconstructs 
the quantum state of light by directly
measuring the planar Radon transform of the Wigner function \cite{Smithey93,Leonhardt97}.
Also, the Husimi Q function \cite{husimi1940}
has been experimentally measured for various systems 
\cite{bohnet2016,haas2014,strobel2014,kanem2005,Eichler11,Agarwal98,bouchard2016}.
We detail how to tomographically  reconstruct a class of finite-dimensional
phase-space representations.

\begin{figure*}[tb]
	\begin{centering}
		\includegraphics{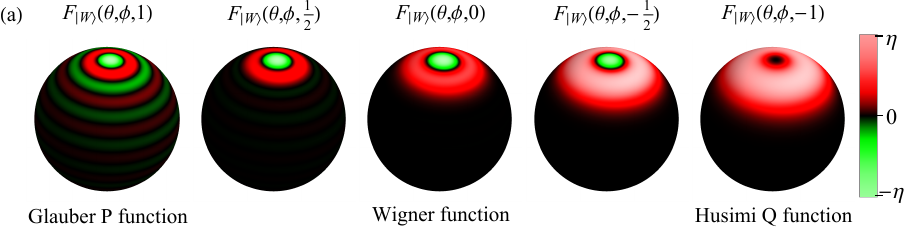}\\[4mm]	
		\includegraphics{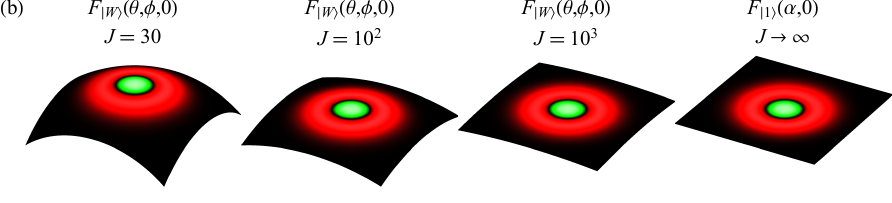}
		\caption{
				(a) $s$-parametrized phase-space representations $F_{|W\rangle}(\theta,\phi,s)$
				for $s\in\{1,1/2,0,-1/2,-1\}$ of a 
				(generalized) W state $|\text{W}\rangle$ for a single spin with $J=10$,
				or equivalently the symmetric Dicke state $| J,J{-}1 \rangle$ of $2J$ indistinguishable qubits
				with a single Majorana vector pointing to the south pole and $2J{-}1$ vectors pointing
				to the north pole.
				A decreasing $s$ (left-to-right) which smears out $F_{|W\rangle}(\theta,\phi,s)$ is
				interpreted as a Gaussian-like convolution.
				Red (dark gray) and green (light gray) represent positive and negative values, respectively. The brightness
				reflects the absolute value of the function relative to its global maximum $\eta$.
				(b)
				Spherical Wigner functions $F_{|W\rangle}(\theta,\phi,0)$ for increasing $J$
				approach their 
				planar counterpart, i.e., the single-photon state $F_{|1\rangle}(\alpha,0)$
				(see Sec.~\ref{summary}).
				Identical coordinate patches with $-1.2 \leq x,y \leq 1.2$ have been used,
				where $x=R\sin\theta \cos\phi$, $y=R\sin\theta \sin\phi$ in the first three
				plots
				and $x=\Re(\alpha)$, $y=\Im(\alpha)$ in the last one.
				(For the plots in (b), methods from \cite{KZG} to
				efficiently approximate phase-space representations
				for large $J$
				have been applied.)
	                 \label{Wstate}
			}
	\end{centering}
\end{figure*}

In this work, we develop a general and unified description of continuous
phase-space representations for quantum states
of a single spin with arbitrary, integer or half-integer spin number $J$
(i.e.\ a qudit with $d=2J{+}1$),
which is simultaneously applicable to 
experimental bosonic systems consisting of  
indistinguishable qubits \cite{Dicke1954,stockton2003,toth2010,lucke2014}.
A single qudit can be identified with
a bosonic system consisting of $2J$ indistinguishable 
qubits: Figure~\ref{Wstate} depicts a quantum state of 
a single qudit (i.e.\ a single spin $J$)  corresponding to
a (generalized) W state \cite{DVC00} (i.e.\ Dicke state)
of $2J$ indistinguishable qubits
(see also Sec.~III~A of \cite{stockton2003}
for an explicit map and Chap.~3.8 of \cite{sakurai1995modern} or \cite{schwinger65}
for links to the second quantization).
In particular, we address the following fundamental open questions
related to finite-dimensional phase-space representations
(e.g., Glauber P, Wigner, and Husimi Q):
(a) How can they be systematically defined to naturally recover the infinite-dimensional case of quantum optics 
in the limit of large $J$?
(b) How can they be transformed into each other?
(c) How can their experimental tomographic approaches be formulated in a unified way?

We present answers to these questions for the full class of (finite-dimensional) $s$-parametrized
phase-space representations with $-1 \leq s \leq 1$.
Our approach relies on 
rotated parity operators and thereby significantly simplifies earlier work 
(such as \cite{Brif98} and particular cases
discussed in \cite{Agarwal81,DowlingAgarwalSchleich}).
It also extends \cite{heiss2000discrete,KdG10,tilma2016,rundle2017,RTD17} 
in the case of single spins (and bosonic systems consisting of  
indistinguishable qubits) to all $s$-parametrized phase spaces.
In addition to a deeper theoretical knowledge connecting
planar and spherical phase spaces, the insights provided here will also guide practitioners
to design innovative experimental schemes, such 
as the tomographic reconstruction of phase-space representations.
Before discussing finite-dimensional quantum states, we first review important properties
of the infinite-dimensional phase spaces from quantum optics.

\section{Summary of infinite-dimensional phase-space representations\label{summary}}
Let us recall the $s$-parametrized 
phase-space distribution 
function (where $-1 \leq s \leq 1$)
\begin{equation}
\label{inifnitedimdefinition}
F_\rho (\Omega,s) = \mathrm{Tr}\,[ \, \rho \, \mathcal{D}(\Omega)  \Pi_s  \mathcal{D}^\dagger(\Omega) ]
\end{equation}
as
the expectation value of the parity operator
$\Pi_s$ (\emph{vide infra})
transformed by the displacement operator
$\mathcal{D}(\Omega)$,
which acts  on coherent states via $\mathcal{D} (\Omega) |0\rangle = |\Omega \rangle$
\cite{Gazeau}, refer also to \cite{cahill1969,moya1993,Leonhardt97,thewignertransform}.
We have developed and discussed the theoretical foundations for the case of infinite dimensions
considered in formula~\eqref{inifnitedimdefinition} in Ref.~\cite{koczor2018}, while building
on earlier work by Grossmann \cite{Grossmann1976} for Wigner functions.
Here, $|0\rangle$ denotes the vacuum state and $\Omega$ 
fully parametrizes a phase space 
with either the variables $p$ and $q$ or
the complex eigenvalues $\alpha$
of the annihilation operator \cite{Glauber1963,Leonhardt97}.

\begin{figure}[tb]
	\begin{centering}
		\includegraphics{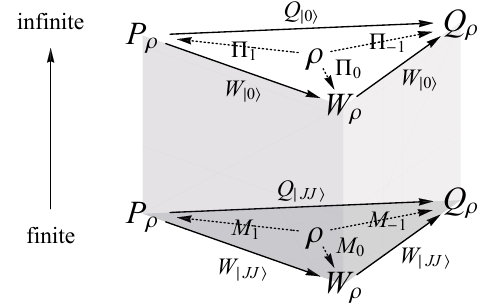}
		\caption{Phase-space representations $W_\rho$, $Q_\rho$,  $P_\rho$
			of \emph{infinite}- or \emph{finite}-dimensional density operators $\rho$
			as expectation value
			of parity operators $\Pi_s$ or $M_s$ [dashed arrows], see
			Eqs.~\eqref{inifnitedimdefinition} or \eqref{PSrepDefinition}.
			Transformed
			by Gaussian smoothing with $W_{|0\rangle}$, $Q_{|0\rangle}$  
			or reversibly with
			$W_{\spinvacuumpure}$, $Q_{\spinvacuumpure}$  [solid arrows], 
			see Eqs.~\eqref{infinitedimensionalswitch} or \eqref{spinswitchrep}.
			\label{CommutativeDiagram}}
	\end{centering}
\end{figure}

Different parity operators $\Pi_s$ lead to different distribution functions
$F_\rho (\Omega,s)$. The Q function $Q_\rho = Q_\rho(\Omega) := F_\rho (\Omega,-1)$
arises from the parity operator $\Pi_{-1}$ whose entries are given by
$[\Pi_{-1}]_{nn} := \delta_{n0}$ \cite{moya1993}
in the number state representation \cite{Leonhardt97}.
Similarly, the Wigner function $W_\rho  := F_\rho (\Omega,0)$
is determined by $[ \Pi_0 ]_{nn}  = 2 (-1)^n$ \cite{moya1993},
which inverts phase-space coordinates via 
$\Pi_0 |\Omega \rangle = |{-} \Omega \rangle$ \cite{thewignertransform}.
The P function $P_\rho := F_\rho (\Omega,1)$ 
is singular for all pure states \cite{cahill1969},
and the entries of its parity operator $\Pi_1$ diverge
in the number-state representation \cite{moya1993}.
The discussed representations 
are considered in the upper part of
Fig.~\ref{CommutativeDiagram}.
An example is given by the vacuum state $|0\rangle$
whose Wigner function $\wnull = 2 e^{-2 |\alpha|^2}$ is a Gaussian distribution.
The respective Q function $\Qnull = e^{- |\alpha|^2}$ is a Gaussian 
of double width and the P function is the two-dimensional
delta function $\Pnull = \delta^{(2)}(\alpha)$. 

We now recollect how to transform
between phase-space representations with Gaussian convolutions \cite{cahill1969,Leonhardt97}.
Two phase-space distribution functions $K(\Omega)$ and $F(\Omega)$
can be combined using their 
convolution \cite{Leonhardt97}
\begin{equation}
\label{infiniteconvolution}
[ K \ast F  ] (\Omega)
=
\int [ \mathcal{D}^{-1}(\Omega)  K(\Omega') ]   F(\Omega') \, \mathrm{d} \Omega',
\end{equation}
which corresponds to a multiplication in the Fourier domain.
Convolution of a distribution function $F_\rho(\Omega,s)$
with the vacuum-state representation $F_{| 0 \rangle}(\Omega,s')$ 
results in the phase-space distribution function
\begin{equation}
\label{infinitedimensionalswitch}
F_\rho(\Omega,s{+}s'{-}1) = F_{| 0 \rangle}(\Omega,s')  \ast F_\rho(\Omega,s)
\end{equation}
of type $s{+}s'{-}1$.
A convolution $\Pnull(\Omega) \ast F(\Omega) = F(\Omega)$ with the P function $\Pnull$
acts as an identity operation, while
a convolution with the Gaussians $\wnull$ or $\Qnull$ 
blurs out $F_\rho(\Omega,s)$. 
This Gaussian smoothing
is widely used in image processing and allows us 
to transform different phase-space representations into each other \cite{Leonhardt97} 
as in the upper part of
Fig.~\ref{CommutativeDiagram}.
For example, the non-negative Q function
$Q_\rho = \wnull \ast W_\rho$ is obtained from the
Wigner function $W_\rho$ by 
convolution with $\wnull$;
the negative regions in $W_\rho$ are therefore bounded
by the variance $1/4$ of 
$\wnull$ \cite{Leonhardt97}.

\section{Phase-space representations for spins\label{main_part}}

\subsection{Definition of phase-space representations for spins\label{subsec_def}}

We establish 
a consistent formalism for $s$-parametrized phase-space
representations ($-1\leq s \leq 1$) for quantum states of single spins,
which in the limit of an increasing spin number $J$ converges
to the just discussed  infinite-dimensional case. 
The continuous phase space
$\Omega := (\theta,\phi)$ 
can be completely parametrized in terms of two Euler angles
of the rotation operator
$\mathcal{R}(\Omega)=\mathcal{R}(\theta,\phi):= e^{i\phi \mathcal{J}_z} e^{i\theta \mathcal{J}_y} $,
where
$\mathcal{J}_z$ and $\mathcal{J}_y$ are components of the 
angular momentum operator \cite{messiah1962}.
The rotation operator $\mathcal{R}(\Omega)$ replaces the displacement operator
$\mathcal{D}(\Omega)$
and maps the spin-up state $\spinvacuumpure$ to 
spin coherent states $| \Omega \rangle = \mathcal{R}(\Omega) \spinvacuumpure$
\cite{perelomov2012,arecchi1972atomic,DowlingAgarwalSchleich,Gazeau}.
This leads to a spherical phase space, 
whose radius is set to $R:=\sqrt{J/(2\pi)}$.

\begin{result} \label{result1}
	For a density operator $\rho$ of a single spin $J$, 
	the $s$-parametrized phase-space
	representation [cf.\ Eq.~\eqref{inifnitedimdefinition}]
	\begin{equation}
	\label{PSrepDefinition}
	F_\rho (\Omega,s) :=  \Tr \,[ \, \rho \, \mathcal{R}(\Omega)  M_s  \mathcal{R}^{\text{\emph{$\dagger$}}}(\Omega) ]
	\end{equation}
	is the expectation
	value of the rotated parity operator
	\begin{equation}
	\label{DefOfspinParityoperators}
	M_s := \tfrac{1}{R} \, \sum_{j=0}^{2J} \sqrt{\tfrac{2j{+}1}{4 \pi}} (\gamma_j)^{-s} \, \T_{j0},
	\end{equation}
	which is a weighted sum of zeroth-order tensor operators. 
\end{result}

In Result~\ref{result1}, the diagonal tensor
operators  
$[\T_{j0}]_{mm'}= 
\delta_{mm'} \sqrt{ (2j{+}1) /(2J{+}1) } \, C_{Jm,j0}^{Jm}$ 
of order zero \cite{Racah42} have been applied
in Eq.~\eqref{DefOfspinParityoperators}, and they can be specified via the Clebsch-Gordan coefficients $C_{Jm,j0}^{Jm}$
\cite{messiah1962} where $j\in \N {\cup} \{0\}$
and $m,m'\in\{-J,\ldots,J\}$. We also use
the coefficients $\gamma_j:=R\, \sqrt{4\pi} (2J)! \, [ (2J{+}j{+}1)! \, (2J{-}j)! \,  ]^{{-}1/2}$.
With increasing spin number $J$, the parity operators
$M_s$ converge to the infinite-dimensional operators $\Pi_s$ in Eq.~\eqref{inifnitedimdefinition},
refer to \cite[Theorem~2.1]{thesis} for a proof, 
while rotations transform into translations along the tangent of a sphere
\cite{amiet2000contracting,arecchi1972atomic,DowlingAgarwalSchleich,klimovgeneralized}.
The phase-space representations in Eq.~\eqref{PSrepDefinition}
fulfill the Stratonovich postulates \cite{stratonovich1956,Agarwal81,vgb89,Brif98,koczor2016};
an $s$-parametrized version is given in Ref.~\cite{Brif98}.
Prior results
\cite{heiss2000discrete,KdG10,tilma2016,rundle2017,RTD17} 
using rotated parity operators
are extended for single spins to all $s$-parametrized phase spaces.
For Wigner functions, our definition conforms to
\cite{RTD17} but differs from Eq.~(8) in \cite{tilma2016}.
The latter can be identified as a linearly shifted Q function $a Q_\rho - b$,
and it relaxes the postulate $\tr(A B) = \int_{S^2}  F_{A}(\Omega,0)  F_{B}(\Omega,0) \, \mathrm{d} \Omega$.
We consider in this work only spherical rotations (even for qudits)
which yield spherical phase spaces,
forgoing general rotations \cite{KdG10,TN2012,tilma2016,rundle2017}.
Generalizations to coupled spins are known in the Wigner case \cite{DROPS,tilma2016,rundle2017};
our methods in \cite{koczor2016} are also applicable.

\begin{figure*}[tb]
	\begin{centering}
		\includegraphics{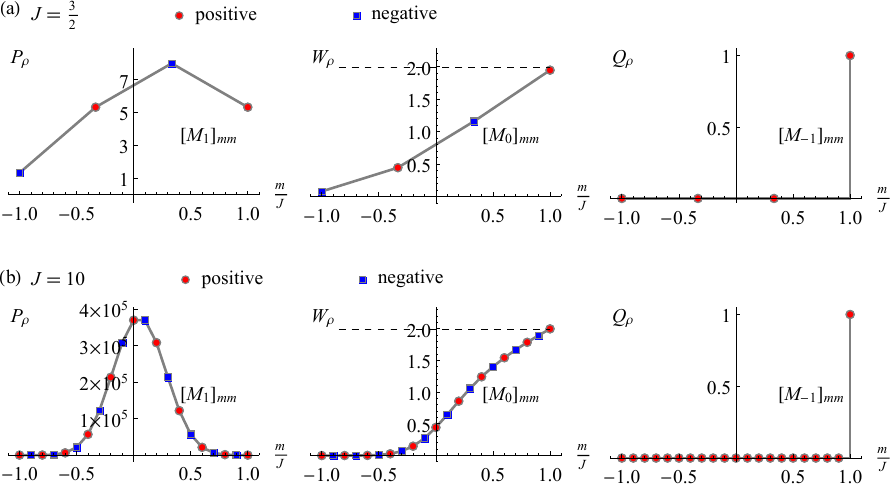}
		\caption{Parity-operator entries $[M_s]_{mm}$
		        corresponding to Eq.~\eqref{DefOfspinParityoperators}		
			[and equivalently the Stern-Gerlach reconstruction weights  in Eq.~\eqref{measurementDef}]
			for a single spin $J$ shown for
			the P function $P_\rho$,
			the Wigner function $W_\rho$, and the Q function $Q_\rho$.
			\label{Measurement_Weights}}
	\end{centering}
\end{figure*}

We further highlight how the approach 
of Result~\ref{result1} connects
to earlier work. An equivalent form of the $s$-parameterized phase-space representation
in Eq.~\eqref{PSrepDefinition}
has been previously determined 
in Eq.~(5.28) of \cite{Brif98} (up to a global factor) as
\begin{gather}\label{phase_old_one}
F_\rho (\Omega,s)
=  \Tr \,[ \, \rho \, \Delta_s(\theta,\phi) ],
\\ \nonumber
\text{with } \; \Delta_s(\theta,\phi)
:= \tfrac{1}{R} \, \sum_{j=0}^{2J} \sum_{m=-j}^{j}  (\gamma_j)^{-s} \, \T_{jm}\, [\Y_{jm}(\theta,\phi)]^*
\end{gather}
using the kernel $\Delta_s(\theta,\phi)$.
Here, $[\Y_{jm}(\theta, \phi)]^*$
denotes the complex conjugate of $\Y_{jm}(\theta, \phi )$.
The work of \cite{Brif97,Brif98} builds on the particular cases
of $s \in \{-1,0,1\}$ obtained in \cite{vgb89}.
Along similar lines, the pioneering work of \cite{Agarwal81} proposed
spherical-harmonics expansions (see Eq.~(3.15) in \cite{Agarwal81})
for  spin phase-space representations 
\begin{equation}\label{phase_old_two}
F^{(\mathit{\Omega})}_\rho (\theta,\phi)
= \sum_{j=0}^{2J} \sum_{m=-j}^{j}   \, c^{(\mathit{\Omega})}_{jm} \, \Y_{jm}(\theta,\phi),
\end{equation}
which are indexed by $\mathit{\Omega}=\mathit{\Omega}_{jm}$ and use the coefficients
$ c^{(\mathit{\Omega})}_{jm}=\Tr \,[ \rho\,  \T_{jm}^\dagger ]/\mathit{\Omega}_{jm}$.
For $s$-parametrized phase spaces, one has 
$\mathit{\Omega}=\mathit{\Omega}_{jm}
= R \gamma_j^s$.
Note that \cite{Agarwal81} established  the explicit form of $\mathit{\Omega}_{jm}$ only for 
Husimi Q functions, i.e., $s=-1$. 
The case of Wigner functions ($s=0$) has been discussed in \cite{DowlingAgarwalSchleich}.
Note that the tensor-operator components $\T_{jm}$ can be explicitly given
as $[\T_{jm}]_{m_1 m_2} = 
\sqrt{(2j{+}1)/(2J{+}1)} \, C^{J m_1}_{J m_2, j m}=(-1)^{J-m_2}\, C^{jm}_{Jm_1,J,-m_2}$
using Clebsch-Gordan coefficients 
and $m_1,m_2 \in \{J,\ldots,-J\}$ \cite{messiah1962,Brif98,BL81,Fano53}.

By using rotated parity operators,
the approach of Result~\ref{result1} 
has important conceptual advantages 
when compared to Eqs.~\eqref{phase_old_one}-\eqref{phase_old_two}.
First, Result~\ref{result1} 
separates the dependence on the parameter $s$ in the parity operator from the 
rotations.
Second, Eq.~\eqref{PSrepDefinition} naturally transforms in the large-spin limit
into the infinite-dimensional case discussed in Eq.~\eqref{inifnitedimdefinition}
by replacing rotations $\mathcal{R}(\Omega)$ with displacements $\mathcal{D}(\Omega)$.
Third, the above mentioned tensor operators and spherical-harmonics decompositions are averted 
and the rotations $\mathcal{R}(\Omega)$ can be efficiently
calculated via the Wigner D-matrix \cite{wigdmatrix1,wigdmatrix2}.
Finally, the particular form given in Result~\ref{result1} enables us to develop
general tomography formulas in Sec.~\ref{tomo}  below.

Particular cases of Result~\ref{result1}
are considered  in the lower part 
of Fig.~\ref{CommutativeDiagram}.
The Q function specifies the expectation value of rotated spin-up states,
where $[M_{-1}]_{mm} := \delta_{mJ}$ (right of Fig.~\ref{Measurement_Weights}),
and its zeros are 
the so-called Majorana vectors
\cite{devi2012majorana,bouchard2016,bjork2015}.
The Wigner function determines the expectation value of the rotated parity operator $M_{0}$.
The matrix entries $[M_0]_{mm}$ are shown in the middle of Fig.~\ref{Measurement_Weights},
highlighting their 
infinite-dimensional limit of $\pm 2$ 
for
$m/J \approx 1$ \cite{amiet2000contracting}. 
The matrix entries  $[M_1]_{mm}$ for the parity operator  of the P function are shown
in the left panel of Fig.~\ref{Measurement_Weights}, including their rapid divergence.

\begin{figure*}[tb]
	\begin{centering}
		\includegraphics{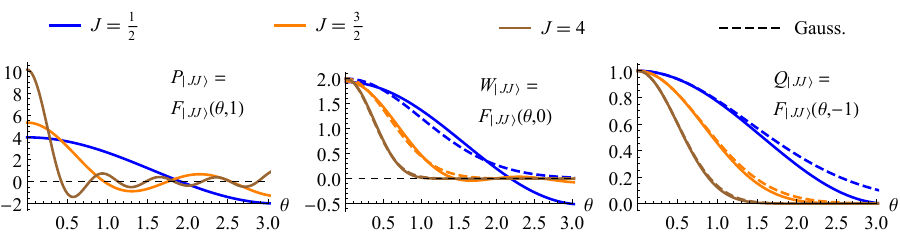}
		\caption{
			Phase-space representations $F_{\spinvacuumpure}(\theta,s)$
			of the spin-up state $\spinvacuumpure$,
			c.f.\ Eq.~\eqref{DefOfKernelFunction}.
			As $J$ increases, 
			$Q_{\spinvacuumpure}$
			and $W_{\spinvacuumpure}$
			rapidly converge to
			the Gaussian distributions $Q_{| 0 \rangle}$ and $W_{| 0 \rangle}$ (dashed line);
			$P_{\spinvacuumpure}$ slowly 
			approaches the delta function $\Pnull = \delta(\Omega)$.
			\label{Gaussian_Approx}}
	\end{centering}
\end{figure*}

Further exploring the infinite-dimensional limit of large $J$,
the phase-space representation
\begin{equation}
\label{DefOfKernelFunction}
F_{\spinvacuumpure} (\theta,s) := \frac{1}{R^2} \, \sum_{j=0}^{2J} 
\sqrt{\tfrac{2j+1}{4 \pi}} (\gamma_j)^{1-s} \, \Y_{j0}(\theta)
\end{equation}
of the spin-up state (i.e.\ the ground state with least uncertainty)
is easily expanded into a weighted sum of axially symmetric spherical harmonics $\Y_{j0}(\theta)$.
The examples $Q_{\spinvacuumpure}$, $W_{\spinvacuumpure}$, 
and $P_{\spinvacuumpure}$ are plotted in Fig.~\ref{Gaussian_Approx}
as functions of the angle $\theta$. 
Even though the Gaussian width of $F_{\spinvacuumpure} (\theta,s)$ 
shrinks in terms of $\theta$ with  increasing $J$, 
$F_{\spinvacuumpure} (\theta,s)$ converges to  
the Gaussian $F_{|0\rangle}(\Omega,s)$
related to the infinite-dimensional vacuum state
if parametrized by the relevant arc length $a:=\theta  R = \theta \sqrt{J/(2\pi)}$
(Fig.~\ref{Wstate}(b) illustrates the sphere-to-plane transition in the infinite-dimensional limit).

For example, $Q_{\spinvacuumpure}$ 
is
equal to the Wigner D-matrix element $|D_{JJ}^J|^2 = \cos{(\theta/2)}^{4J}$,
and  it converges rapidly with increasing $J$ to the Gaussian 
$Q_{|0 \rangle}(\alpha)= e^{- |\alpha|^2}=e^{- a^2 \pi}=e^{- J \theta^2 / 2}$
using the phase-space coordinate $\alpha = \sqrt{\pi} a e^{-i\phi}$
\cite{perelomov2012,arecchi1972atomic}.
Similarly, $W_{\spinvacuumpure}$
rapidly converges to the normalized Gaussian $\wnull =2  e^{-2 |\alpha|^2} = 2e^{- 2 a^2 \pi} =
2  e^{- J \theta^2}$ of the vacuum state.
The P function $P_{\spinvacuumpure}(\theta) := \tilde{\delta} (\Omega)$
is the spherical $sinc$ function, i.e., a truncated version of the spherical
delta function $\delta(\Omega):=\delta(\theta) \delta(\phi)/ \sin{\theta}$
(where the tilde projects onto the physical subspace of spherical harmonics
with rank $j \leq 2J$ \cite{giraud2008}),
which by definition approaches the delta function
in the large-spin limit 
$\delta(\Omega):=\sum_{j=0}^{\infty} \sqrt{(2j{+}1)/(4 \pi)} \, \Y_{j,0}(\Omega)$.
Qualitative similarities between certain finite- and infinite-dimensional
Wigner functions were already highlighted in \cite{DowlingAgarwalSchleich}.
But this connection is clarified in our formulation by emphasizing
the large-spin convergence for all of the $s$-parametrized phase spaces,
refer to \cite[Theorem~2.1]{thesis} for a proof.

\subsection{Spherical convolution\label{spherical_conv}}

To translate 
between the different spherical phase-space representations
in the lower part of Fig.~\ref{CommutativeDiagram}
(which can be done reversibly assuming arbitrary precision), we define
the convolution  [cf.\ Eq.~\eqref{infiniteconvolution}]
\begin{equation}
\label{ShericalConv}
[ K \ast F  ] (\Omega)
:= 
\int_{S^2} [ \mathcal{R}^{-1}(\Omega)  K(\Omega') ]   F(\Omega') \, \mathrm{d} \Omega'
\end{equation}
via a spherical integration where 
$\mathrm{d} \Omega' = R^2 \sin{\theta'} \mathrm{d} \theta' \, \mathrm{d} \phi'$. 
First, the kernel function $K(\Omega')$ is rotated by $\mathcal{R}^{-1}(\Omega)$
to $K(\Omega'{-}\Omega)$, which 
is then projected onto the distribution function $F(\Omega')$ via a spherical integral.
The kernel function $K(\Omega')$ has to be axially symmetric due to
the so-called Funk-Hecke theorem \cite{groemer1996,kennedy2013book}.
The spherical convolution
is a multiplication in the spherical-harmonics domain, and substituting
spherical harmonics into Eq.~\eqref{ShericalConv} yields 
$\Y_{j'0} \ast \Y_{jm} = R^2 \sqrt{4\pi/(2j{+}1)} \, \Y_{jm} \, \delta_{jj'}$.
This allows us to transform between different spherical phase-space representations:
\begin{result} \label{result2}
	The convolution of a phase-space distribution function $F_\rho(\Omega,s)$
	with  the phase-space representation $F_{\spinvacuumpure}(\Omega,s')$
	of the
	spin-up state 
	results in
	a type-$(s{+}s'{-}1)$ distribution function [cf.\  Eq.~\eqref{infinitedimensionalswitch}]
	\begin{equation}
	\label{spinswitchrep}
	F_\rho(\Omega,s{+}s'{-}1) = F_{\spinvacuumpure} (\theta,s') \ast F_\rho(\Omega,s).
	\end{equation}
\end{result}
The pioneering work of \cite{Agarwal81} proposed
spin phase-space representations in the form of
spherical-harmonics expansions [refer to Eq.~\eqref{phase_old_two}]
and defined their relations using integral transformation kernels
(see (3.19) in \cite{Agarwal81}). Result~\ref{result2} clarifies that these relations are in fact
spherical convolutions, in complete analogy with the infinite-dimensional case 
considered in quantum
optics. The general form of Eq.~\eqref{spinswitchrep}
has not been formally described in the literature before.
Some convolution properties were detailed for \emph{discrete}, planar phase spaces
in \cite{buvzek1995,opatrny1995}.
We want to also stress that spherical convolutions have
efficient implementations \cite{beamdeconv,fastconv}.

In the infinite-dimensional limit of an increasing spin number $J$, Eq.~\eqref{spinswitchrep}
turns into Eq.~\eqref{infinitedimensionalswitch}.
We emphasize that the convolution transformation in Eq.~\eqref{spinswitchrep} is reversible
(assuming arbitrary precision) for
general  parameters $s,s'\in\mathbb{R}$ (as the coefficients $\gamma_j$ in Eq.~\eqref{DefOfKernelFunction}
are non-zero).
Also, a convolution 
$P_{\spinvacuumpure}(\theta) \ast F(\Omega,s) = F(\Omega,s)$ 
with the P function $P_{\spinvacuumpure}(\theta)$
acts as an identity operation, just as in the infinite-dimensional case. 
The Wigner function $W_\rho$ can be transformed into
the non-negative Q function
$Q_{\rho}= W_{\spinvacuumpure} \ast W_\rho$
by Gaussian-like smoothing, cf.\ Fig.~\ref{Wstate}. Consequently,
the negative regions of $W_\rho$ 
are bounded by the variance $\propto1/4$ of $W_{\spinvacuumpure}$,
similar as for infinite-dimensional phase spaces.
Result~\ref{result2} completes 
our characterization of how to transform between spherical phase-space 
representations as illustrated in Fig.~\ref{CommutativeDiagram}.

\addtocounter{footnote}{1}
\footnotetext[\value{footnote}]{
The 
depicted random pure state vector is 
approximately given by $(
0.06 + i0.02,\,
-0.21 - i0.19,\,
0.04 + i0.27,\,
0.15 - i0.11,\, 
0.28 - i0.28,\,
-0.33 - i0.25,\,
0.04 - i0.44,\,
-0.21 - i0.24,\,
-0.43 + i0.00
)^T$.}
\newcounter{footcombined}
\setcounter{footcombined}{\value{footnote}}
\begin{figure*}[tb]
	\begin{centering}
		\includegraphics{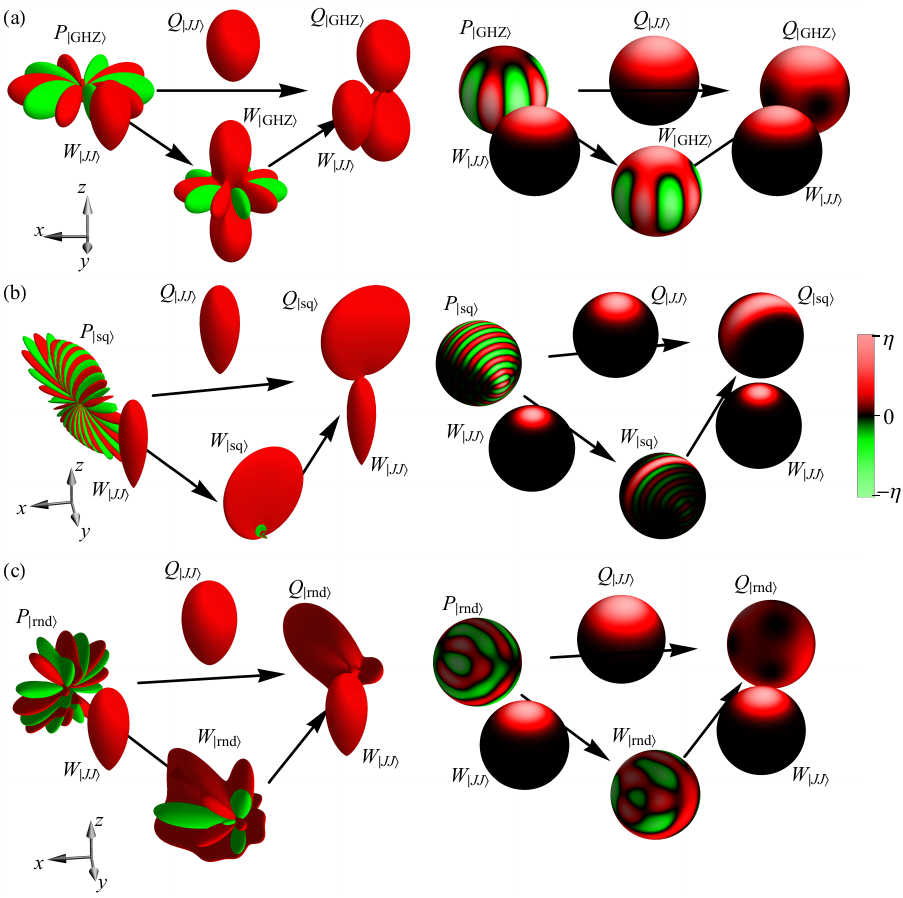}
		\caption{ 
			P, Wigner, and Q functions with their corresponding convolution
			kernels
			for (a) a quantum state of a spin $J=5/2$
			corresponding to the
			GHZ state $|\mathrm{GHZ}\rangle$ of $2J$ indistinguishable qubits,
			(b) a squeezed state $|\mathrm{sq}\rangle$ of a spin $J=10$, and
			(c) a random state $|\mathrm{rnd}\rangle$ of a spin $J=4$
			\cite{Note\thefootcombined}
			(see Sec.~\ref{exampleSec}). 
			Red (dark gray) and green (light gray) represent positive
			and negative values, respectively.
			The absolute values of the spherical function 
			relative to its global maximum $\eta$ is given by
			the plotted surface (left) or the brightness (right), where each variant 
			highlights different properties of the plotted functions.			
                          \label{examplesFig}
		}
	\end{centering}
\end{figure*}

\subsection{Examples of phase-space functions \label{exampleSec}}

Figure~\ref{examplesFig} depicts phase-space representations of typical finite-dimensional
quantum states.
The P, Wigner, and Q functions are shown in a triangular arrangement along with
their corresponding convolution kernels, which generate the 
spherical convolutions from Eq.~\eqref{spinswitchrep} between edges of the triangle.

In  Fig.~\ref{examplesFig}(a), we consider the quantum state of a single spin $J$ 
corresponding to the
$2J$-qubit GHZ state $|\text{GHZ}\rangle 
=(| 0 \rangle^{\otimes 2J} + | 1 \rangle^{\otimes 2J})/\sqrt{2}=
( | J\!J \rangle + | J\!,{-}J \rangle ) /\sqrt{2}$
consisting of a quantum superposition of the two symmetric Dicke states given by  the spin-up and spin-down state
(which can be identified with a $2J$-photon NOON state).
This GHZ state factorizes 
up to permutations into a product of
its Majorana vectors $ \bigotimes_k |v_k \rangle$ \cite{devi2012majorana,bjork2015},
where $|v_k \rangle$ is a single-qubit state with Bloch vector $v_k$.
These Majorana vectors correspond to zeros of the Q function and 
point to the edges of a regular $n$-gon, see $Q_{|\mathrm{GHZ}\rangle}$ in Fig.~\ref{examplesFig}(a).
The zeros of the Q funtion can (e.g.) be determined by spherically convolving 
the Wigner function with  the convolution kernel $W_{\spinvacuumpure}$
(cf.\ Sec.~\ref{spherical_conv}), and the negative (green) lobes 
of $P_{|\mathrm{GHZ}\rangle}$
and $W_{|\mathrm{GHZ}\rangle}$ in Fig.~\ref{examplesFig}(a)
identify the direction of the Majorana vectors.
The Q function largely resembles the classical superposition of a spin-up and a spin-down state, 
but has a five-fold symmetry.

Figure~\ref{examplesFig}(b) shows phase-space plots of the
squeezed state $\exp[- i \theta \, \mathcal{I}_y^2/2]\spinvacuumpure$
with squeezing angle $\theta := 0.3$ for
a single spin with spin number $J=10$, where the state is squeezed along the 
$y$ axis \cite{ma2011quantum}.
A random pure state of a single spin with spin number $J=4$ 
is depicted in Fig.~\ref{examplesFig}(c).

\section{Tomography\label{tomo}}

\subsection{Pointwise tomography of phase-space functions\label{sec:tomo}}

We detail
how phase-space representations 
are recovered from Stern-Gerlach experiments
assuming that a chosen 
density operator $\rho$ can be prepared identically and repeatedly.
In a single Stern-Gerlach experiment,
one detects the density matrix $\rho$ in a projection eigenstate according to a reference 
frame rotated by $\Omega$ (i.e., by rotating the measurement device or inversely rotating $\rho$).
For repeated Stern-Gerlach experiments,
measured frequencies converge to the  
Stern-Gerlach probabilities
\begin{equation}\label{rotated_prob}
p_m (\Omega)= 
\langle Jm | \mathcal{R}^{\dagger}(\Omega)  \rho  \mathcal{R}(\Omega)  |Jm \rangle.
\end{equation}
In the limiting case of infinite-dimensional parity operators 
\cite{koczor2018,thesis,amiet2000contracting}, this is known
in quantum optics as the `direct measurement' technique
\cite{deleglise2008,Lutterbach97,Bertet02,Banaszek99}.
Here, we have the finite-dimensional equivalent:
\begin{result} \label{result3}
	The phase-space representations 
	\begin{equation}
	\label{measurementDef}
	F_\rho (\Omega,s) = \sum_{m=-J}^J [M_s]_{mm} \; p_m (\Omega)
	\end{equation}
	of a $(2J{+}1)$-dimensional quantum state $\rho$ 
	are directly determined for each phase-space
	point $\Omega$ by the  
	probability distributions $p_m (\Omega)$ of repeated
	Stern-Gerlach experiments, see Eq.~\eqref{rotated_prob}. 
	The weights $[M_s]_{mm}$
	are given by the parity operator
	from Eq.~\eqref{DefOfspinParityoperators}.
\end{result}

The pointwise tomography of Result~\ref{result3}
has not been described in this generality before.
We discuss different cases of the parameter $s$ by referring to the examples
of phase-space functions in
Fig.~\ref{examplesFig}.
In particular, the P functions $P_{|\mathrm{GHZ}\rangle}$, $P_{|\mathrm{sq}\rangle}$,
and $P_{|\mathrm{rnd}\rangle}$ in Fig.~\ref{examplesFig}
show considerable detail, while mostly utilizing probabilities $p_m(\Omega)$ of
small $\abs{m}$ (cf.\ Fig.~\ref{Measurement_Weights}).
The Wigner functions $W_{|\mathrm{GHZ}\rangle}$, $W_{|\mathrm{sq}\rangle}$,
and $W_{|\mathrm{rnd}\rangle}$ in Fig.~\ref{examplesFig}
require all $2J{+}1$ Stern-Gerlach probabilities $p_m(\Omega)$
\cite{Schmied2011,treutlein2010,mcconnell2015,Manko97,rundle2017}
and show fewer detail consistent with being smoothed versions of
the corresponding P functions. 
Finally, the Q functions show little detail due to a second
Gaussian smoothing
(yet low-rank contributions would still be recognizable
\cite{lucke2014,bohnet2016,haas2014,strobel2014,Schmied2011})
and are fixed by the probability
$p_J(\Omega)$ of the spin-up state \cite{Agarwal98}.
Certain features of our tomography approach such as the 
weights in Eq.~\eqref{measurementDef}
are invariant under slight variations of 
a sufficiently large spin number $J$, and this
might be useful in atomic ensembles \cite{mcconnell2015,haas2014},
Bose-Einstein condensates \cite{treutlein2010,Schmied2011,anderson1995,ho1998,stenger1999,lin2011},
or trapped ions \cite{leibfried2005,bohnet2016,monz2011}.

We detail how Result~\ref{result3} is applied in the estimation of the
$s$-parametrized phase-space function $F_\rho (\Omega,s)$ of a
quantum state $\rho$ at a single phase-space point $\Omega=(\theta,\phi)$: 
the quantum state is rotated according to the angles $(\theta,\phi)$,
a projective Stern-Gerlach measurement is performed,
the measured eigenstate $m$ is recorded,
and the whole procedure is repeated $\nproj$ times.
Then the probabilities $p_m(\theta,\phi)$ can be estimated
from the relative frequencies $N_m / \nproj$ of the eigenstates,
where the eigenstate $m$ has been recorded $N_m$ times during the 
measurements.
This enables the reconstruction of a phase-space function
at the phase-space point $(\theta,\phi)$
as a linear combination of the estimated
probabilities in Eq.~\eqref{measurementDef}, where the
weights $[M_s]_{mm}$ are illustrated in Fig.~\ref{Measurement_Weights}.
The Wigner-function tomography for a random ensemble of $N_\rho=2200$ spin-$5/2$
states (which are distributed with respect to the Hilbert-Schmidt distance
\cite{zyczkowski2011}) has been simulated with $\nproj=10^2$, $10^3$, and $10^4$ repetitions
for the phase-space point $(\theta,\phi)=(0,0)$ and for each reconstructed random state.
Figure~\ref{reconstructions} (a) shows the reconstruction errors which follow
an empirical Gaussian (i.e.\ normal) distribution. The standard deviation empirically 
scales with $\nproj^{-1/2}$ and therefore
vanishes as the number $\nproj$ of repetitions increases. We now apply the pointwise
tomography first for multiple phase-space points and then to
obtain a full tomography of a phase-space function.

\subsection{Pointwise tomography for multiple phase-space points\label{full_point_wise}}

The pointwise tomography of Result~\ref{result3} can be easily repeated
for multiple phase-space points. This enables an \emph{approximate} pointwise reconstruction of 
phase-space functions: the approximation improves
as the number of phase-space points increases.
Figure~\ref{reconstructions} (b) shows the average error of pointwisely reconstructed phase-space
functions, and this average error reduces as the number of Stern-Gerlach measurements increases.
However, this approximation has a notably 
discrete flavor as it only recovers a phase-space function at the chosen phase-space points and not between them.
In Sec.~\ref{simulations} below, we detail a measurement strategy that relies on a \emph{finite}
number of phase-space points (together with enough Stern-Gerlach repetitions $N_r$) 
in order to recover the full phase-space function as a linear combination
of spherical harmonics.

\subsection{Full tomography of phase-space functions\label{simulations}}

\begin{figure*}[tb]
	\begin{centering}
		\includegraphics{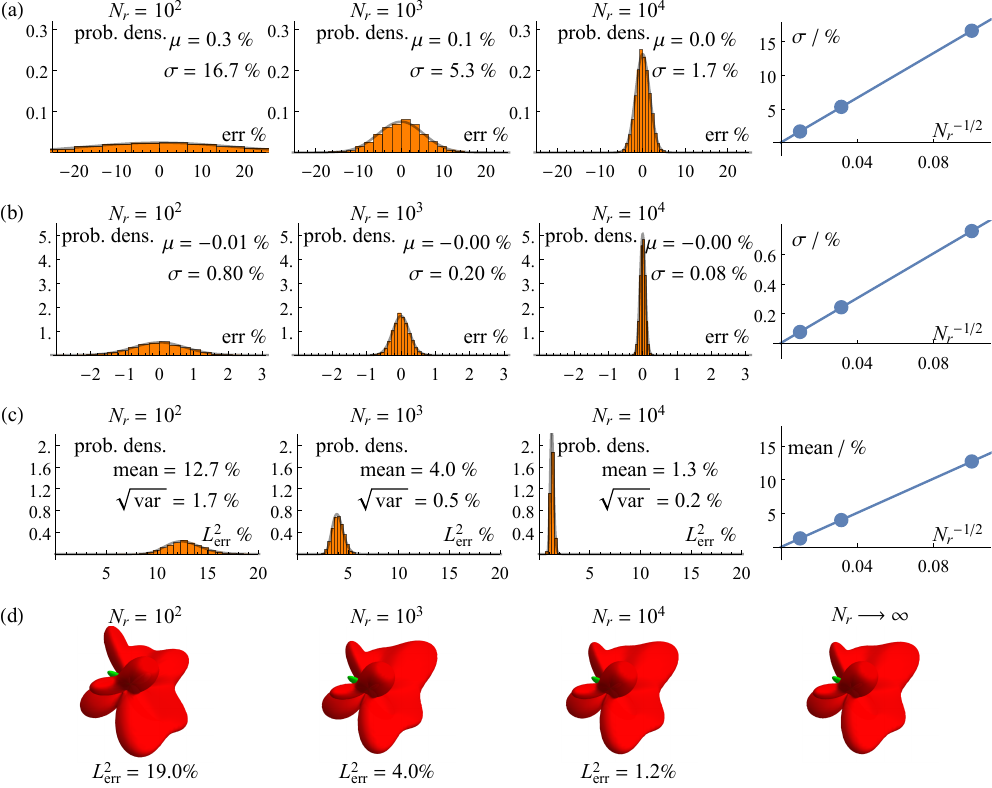}
		\caption{
				Simulated tomography of Wigner functions $W_\rho(\theta,\phi)$ 
				for a random ensemble of $N_\rho = 2200$
				spin-$5/2$ states $\rho$:
				(a)  Pointwise tomography at the phase-space point 
				$(\theta,\phi)=(0,0)$ as discussed in Sec.~\ref{sec:tomo}.
				The relative reconstruction errors (relative to the global maximum of the 
				ideal phase-space function) empirically follow a Gaussian distribution. 
				Its mean $\mu$ and standard deviation $\sigma$ 				
				are obtained from a fitted Gaussian distribution. And $\sigma$
				empirically scales as $\nproj^{-1/2}$ 
				with the number $\nproj$ of repetitions.
				(b)  Pointwise tomographies from (a) evaluated at $22^2 = 484$ phase-space points 
				(as discussed in Sec.~\ref{full_point_wise}) for points $(\theta_k,\phi_q)$ from
				an equiangular grid (refer to Sec.~\ref{simulations}).
				The relative reconstruction errors (relative to the global maximum of the 
					ideal phase-space function) are averaged over the grid points $(\theta_k,\phi_q)$.
					The mean $\mu$ and standard deviation $\sigma$ 	
					of the average error			
					are obtained from a fitted Gaussian distribution; $\sigma$
					empirically scales as $\nproj^{-1/2}$.
				(c) Full tomography as discussed in Sec.~\ref{simulations} using an equiangular grid
				of $22^2 = 484$ phase-space points.
				The relative $L^2$-norm errors 
				(relative to the global maximum of the ideal phase-space function)
				empirically follow a log-normal distribution.
				The mean and standard deviation is determined from a fitted log-normal distribution 
				and the mean also
				scales as $\nproj^{-1/2}$.
				(d) Examples of reconstructed Wigner functions from (c)
				with their relative $L^2$-norm errors. 
			\label{reconstructions}
		}
	\end{centering}
\end{figure*}

The full tomography of a phase-space function
relies on multiple pointwise tomographies.
Assuming that enough repetitions $\nproj$ of
the Stern-Gerlach measurements are performed
for each phase-space point,
the corresponding spherical-harmonics coefficients can then be 
obtained
from pointwise tomographies for 
a \emph{finite} number of phase-space points \cite{driscoll94,kennedy2013book}.
One straight-forward method to determine the spherical-harmonics decomposition
of a spin-$J$ function is by performing pointwise tomographies via Result~\ref{result3} on an equiangular grid of
at least $(4J{+}2)^2$ phase-space points (or combinations of rotation angles).
(This does \emph{not} imply a general lower bound and other measurement 
strategies might be able to use fewer than $(4J{+}2)^2$ phase-space points.)
In this case, one can apply the sampling technique described in
\cite[Theorem~3]{driscoll94} and
\cite[Theorem~7.1]{kennedy2013book}, which
determines a phase-space function as a linear combination of 
spherical harmonics. The equiangular grid is given by 
at least $(4J{+}2)^2$ combinations of rotation angles $\theta_k=(\pi k)/{N_p}$
and $\phi_q=(2\pi q)/{N_p}$ for $k,q \in \{ 0,\ldots,N_p-1\}$
and $N_p\geq  4J{+}2$.
\begin{result} \label{result4}
The complete phase-space function $F_\rho (\Omega,s) = \sum_{j=0}^{2J} \sum_{m=-j}^{j}   \, c_{jm} \, \Y_{jm}(\Omega)$
is determined by its spherical-harmonics expansion coefficients $c_{jm}$ which
are computed from
phase-space reconstructions $\tilde{F}_\rho (\theta_k, \phi_q ,s)$ 
at the phase-space points (or angles) $(\theta_k, \phi_q)$
as
\begin{equation*}
c_{jm} = \frac{2\pi \sqrt{2}}{N_p} \sum_{k=0}^{N_p-1} \sum_{q=0}^{N_p-1}
\alpha^{(N_p)}_k \tilde{F}_\rho (\theta_k, \phi_q, s)\,  [\Y_{jm}(\theta_k, \phi_q )]^*.
\end{equation*}
\end{result}
A closed formula for the real coefficients $\alpha^{(N_p)}_k$ can be
found in \cite{driscoll94,kennedy2013book}.
Increasing the number $N_p$ beyond its lower bound $4J{+}2$ might help to reduce errors
due to experimental imperfections in precisely setting the
rotation angles. Note that  the pointwise reconstructions
$\tilde{F}_\rho (\theta_k, \phi_q, s)$ are usually susceptible to shot noise
(due to the finite number $\nproj$ of Stern-Gerlach repetitions) 
and this also affects the full tomography of the phase-space function.
In Fig.~\ref{reconstructions} (c), the full tomography using Result~\ref{result4}
is simulated for a random ensemble of $N_\rho=2200$ spin-$5/2$ quantum states. 
The reconstruction error is given as the relative
$L^2$-norm difference between the reconstructed and the ideal phase-space
functions (relative to the ideal one)
and it empirically follows a log-normal distribution. The mean of the reconstruction 
error empirically scales as $\nproj^{-1/2}$
and vanishes as the number $\nproj$ of measurements increases.

\begin{figure*}[tb]
	\begin{centering}
		\includegraphics{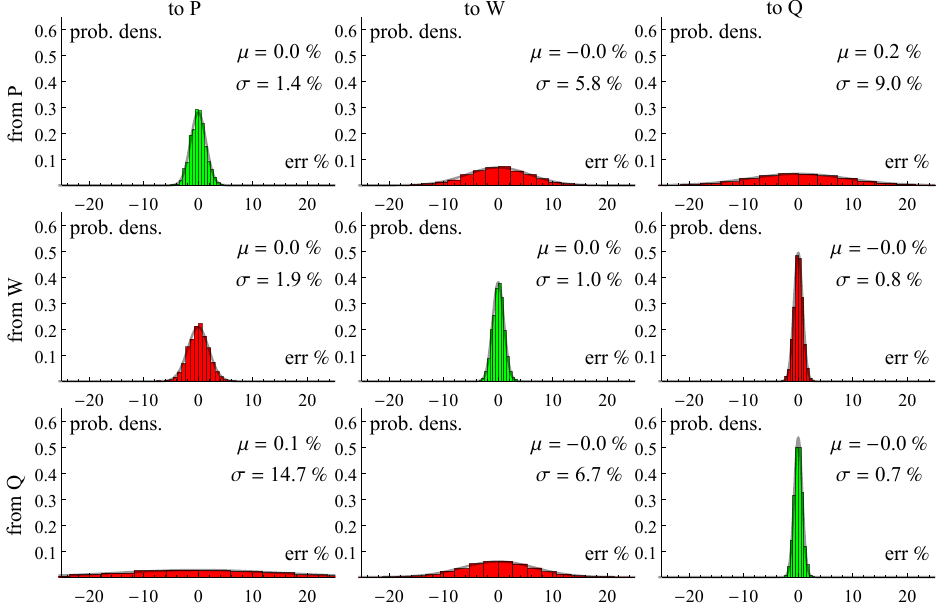}
		\caption{
Relative reconstruction errors 
(relative to the global maximum of the ideal phase-space function)
of
simulated full tomographies of P, W, and Q functions
evaluated at the phase-space point $(\theta,\phi)=(0,0)$
for a random ensemble of $N_\rho=2200$ spin-$5/2$ states
using $\nproj=1000$ Stern-Gerlach repetitions.
This is similar as discussed in Sec.~\ref{simulations} but
the reconstruction errors are only evaluated at $(\theta,\phi)=(0,0)$.
The directly reconstructed phase-space functions (cf.\ Result~\ref{result4})
for the green
histograms on the diagonal
are in a second step transformed 
with a spherical convolution (cf.\ Result~\ref{result2})
to 
P, W, and Q functions for the red, off-diagonal histograms.
The direct reconstruction is usually preferable. 
The mean $\mu$ and standard deviation $\sigma$ 				
are obtained from a fitted Gaussian distribution.
 \label{3x3histograms}}
	\end{centering}
\end{figure*}

Obviously, Result~\ref{result4} describes only one of many 
measurement strategies that can be envisioned by starting from Result~\ref{result3}.
In particular, Result~\ref{result4} uses an equiangular grid 
and results in a concentration of sampling points at the poles.
More isotropic measurement strategies can rely on (e.g.) Lebedev grids~\cite{leb75, leb76, leb99}.
A more detailed and thorough discussion of suitable measurement strategies is left to future research.
In the remaining parts of Sec.~\ref{tomo}, we discuss certain drawbacks 
of combining a tomography 
with a spherical convolution as well as various connections to related work.
Finally, we close this section with a discussion in Sec.~\ref{tomo_discussion}.

\begin{figure*}[tb]
	\begin{centering}
		\includegraphics{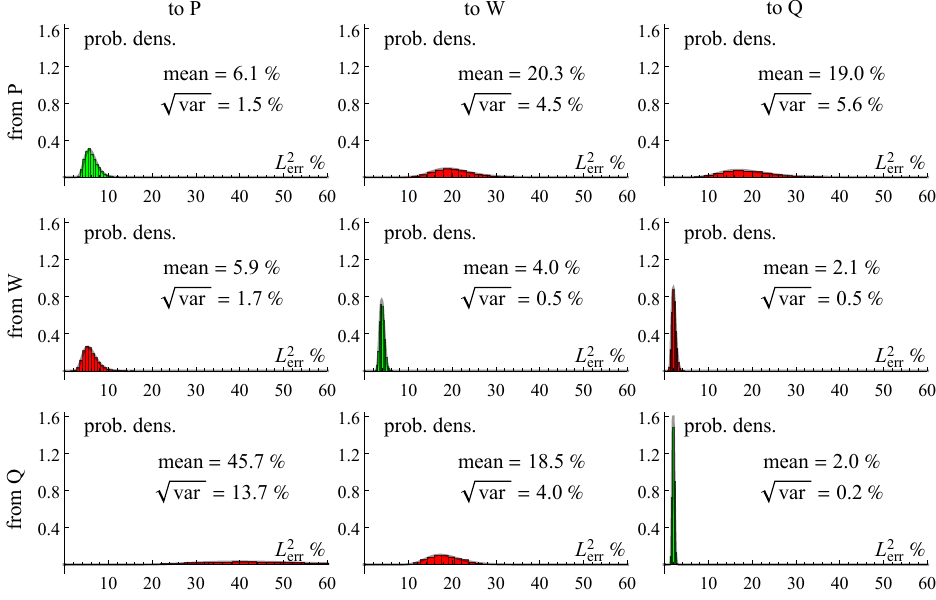}
		\caption{
			Similar as in Fig.~\ref{3x3histograms}, 
			the relative reconstruction errors
			are given here, however, for the full phase-space function as relative $L^2$ errors
			(and not only for a single phase-space point).
			The directly reconstructed phase-space functions for the green
			histograms on the diagonal are usually preferable to 
			the red, off-diagonal histograms that are obtained 
			from the green ones by applying an additional 
			spherical convolution (see Result~\ref{result2}).
			\label{3x3histogramsl2}
		}
	\end{centering}
\end{figure*}

\subsection{Drawbacks of combining a tomography with a spherical convolution\label{pitfalls}}
A reduced number $\nproj$ of Stern-Gerlach repetitions
might lead to a substantial error when one
transforms a reconstructed phase-space function to
a different member
of the
$s$-parametrized class 
of phase-space functions using a spherical convolution (see Result~\ref{result2}).
Figure~\ref{3x3histograms} details this effect
for simulated full tomographies (see Sec.~\ref{simulations}) 
of P, W, and Q functions evaluated at the phase-space point $(\theta,\phi)=(0,0)$.
The stated relative errors are given by the difference between 
the simulated full reconstruction and the ideal phase-space
function (relative to the global maximum
of the ideal one).
First, Result~\ref{result4} is used for a full
tomography of $N_\rho=2200$ random spin-$5/2$ states, where
$\nproj=1000$ Stern-Gerlach repetitions are considered.
This results in the
green histograms on the diagonal of Fig.~\ref{3x3histograms}.
Second, the reconstructed complete phase-space functions are transformed to 
P, W, and Q functions by applying Result~\ref{result2}. One obtains
the red, off-diagonal histograms in Fig.~\ref{3x3histograms}.

Similarly, Fig.~\ref{3x3histogramsl2} considers the
full tomography (see Sec.~\ref{simulations})  and shows
simulated histograms for the relative $L^2$-norm errors between the ideal and
the reconstructed phase-space functions (relative to the $L^2$ norm of the
ideal one). The red, off-diagonal parts for both 
Figure~\ref{3x3histograms} and \ref{3x3histogramsl2}
highlight that one should usually avoid an indirect approach that combines a tomography 
with a spherical convolution from Result~\ref{result2}, at least
for a reduced number $\nproj$ of
Stern-Gerlach repetitions. A direct tomography of 
the desired class of $s$-parametrized phase-space 
function using Result~\ref{result3} or Result~\ref{result4} is preferable.
This highlights that not all reconstruction strategies are equally
advisable under significant errors, even though
the transformations in
Result~\ref{result2} are reversible if one neglects errors.
We have limited our discussion to errors which are a consequence 
of having only a finite number of Stern-Gerlach repetitions
at each phase-space point.

\begin{figure*}[tb]
	\begin{centering}
		\includegraphics{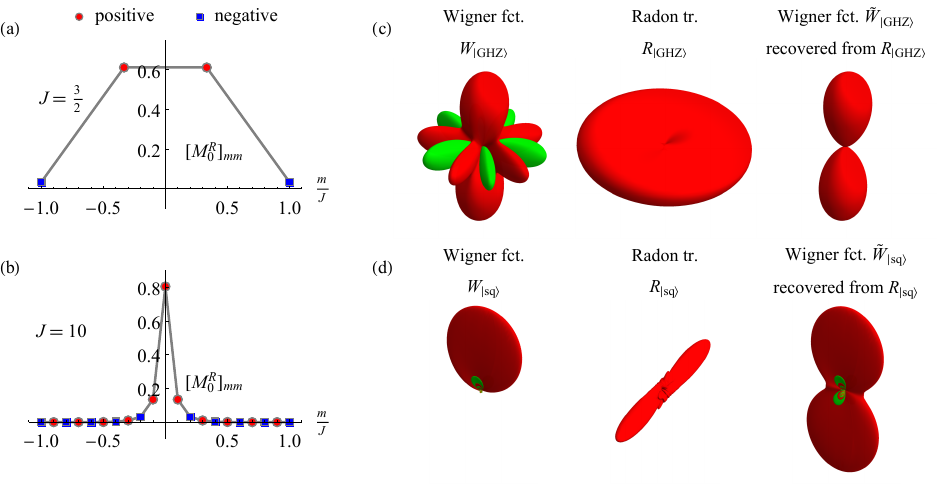}
		\caption{(a)-(b)
			Stern-Gerlach reconstruction weights $[M^R_0]_{mm}$ in Eq.~\eqref{RadonParityOperator}
			for the Radon transform of a Wigner function applicable to a single spin $J$ 
			(cf.\ Fig.~\ref{Measurement_Weights});
			(c)-(d) Wigner function (cf.\ Fig.~\ref{examplesFig}), its Radon transform, and its point-symmetric part
			reconstructed by inverse Radon transformation: 
			(c) quantum state of a single spin 
			with $J=5/2$ corresponding to the
			GHZ state of $2J$ qubits (cf.\ Fig.~\ref{examplesFig}(a)),
			(d) squeezed state $|\mathrm{sq}\rangle$ of a single spin with $J=10$
			(cf.\ Fig.~\ref{examplesFig}(b))
			approximately localized on the upper hemisphere of its Wigner function $W_{|\mathrm{sq}\rangle}$, 
			the corresponding Radon transform 
			$R_{|\mathrm{sq}\rangle}$
			is highly localized around the equator
			and can be reconstructed using few measurements.
			Red (dark gray) and green (light gray) 
			represent positive and negative values, respectively.
			\label{RadonPlots}
		}
	\end{centering}
\end{figure*}

\subsection{Related Experimental Work\label{subsec_related}}

Similar tomography approaches for the reconstruction of phase-space functions
that emphasize rotational symmetries of finite-dimensional quantum systems and
rely on rotated parity operators (as in Result~\ref{result3}) have been 
experimentally validated in the literature \cite{rundle2017,leiner17,Leiner18}.
In \cite{rundle2017}, Stern-Gerlach measurements have been performed
in order to determine the probabilities of finding a quantum system in rotated basis states,
and this allowed them to experimentally recover a particular type of a multi-spin phase-space function
that arises from products of single-spin phase-space functions
(refer to \cite{tilma2016} as discussed in Sec.~\ref{subsec_def}).
The nuclear magnetic resonance experiments in \cite{leiner17}
did not rely on Stern-Gerlach measurements, but directly
measured the overlaps between the mixed quantum state and
rotated axial tensor operators, where generalized multi-spin 
Wigner functions \cite{DROPS} have been experimentally reconstructed  
without first recovering the density matrix. The approach
of \cite{leiner17} has been recently also applied to
the experimental reconstruction of propagators and quantum gates \cite{Leiner18}.
These experiments highlight the convenience of
incorporating rotations directly in the tomography scheme
as we have done in Result~\ref{result3} for the whole class
of $s$-parametrized phase-space functions,
which includes the Glauber P, Wigner, and Husimi Q function.

Let us also compare our work to the
`filtered backprojection' technique in Sec.~2 of \cite{Schmied2011}
(which differs from the spherical Radon approach in Sec.~\ref{subsec_sph_radon}
also discussed in \cite{Schmied2011}):
it relies on the experiments in \cite{treutlein2010}
and recovers a Wigner function from a finite number
$N$ of Stern-Gerlach measurements
(each performed in a rotated reference frame $\Omega_n$):
The Wigner functions $W_{| m_n \rangle}$ 
of the projection eigenstates $|m_n\rangle$
are inversely rotated and summed up
as $\sum_{n=1}^{N} c_n\, \mathcal{R}^{-1}(\Omega_n) [ W_{| m_n \rangle}]$.
A subsequent spherical convolution with a filter function 
reconstructs the Wigner function in \cite{Schmied2011}, 
which is in the limit of infinite and evenly distributed measurements
agrees with the general Result~\ref{result3}. In addition, 
Result~\ref{result3} does not rely on a spherical convolution
and enables diverse reconstruction strategies
as the distribution function $F_\rho (\Omega,s)$
can be independently determined for each phase-space point $\Omega$.

Our comparison to related experimental work
clearly shows the feasibility of our tomography scheme and
the use of rotated parity operators appropriately reflects 
the rotational symmetries of finite-dimensional quantum systems. 
Consequently, we believe that our tomography scheme will
beneficial for a large class of experimental scenarios.

\subsection{Comparison to the spherical Radon approach\label{subsec_sph_radon}}

We also relate Result~\ref{result3} to 
optical homodyne tomography
\cite{Smithey93,Leonhardt97,PR04} (cf.\ Eq.~(6.12) in \cite{Brif98})
and especially to the finite-dimensional case as discussed in \cite{Schmied2011}.
The planar Radon transformation of a Wigner function
is replaced in finite dimensions with the 
spherical Radon transformation,
which is the integral along the great circle orthogonal to the vector
pointing to a phase-space point $\Omega$ \cite{groemer1996}.
Refer to Fig.~\ref{RadonPlots} for plots of the Radon transforms 
$R_{|\mathrm{GHZ}\rangle}$  and $R_{|\mathrm{sq}\rangle}$ 
of Wigner functions for
a GHZ state and a squeezed state, respectively.
The Radon transforms of Wigner functions can be directly obtained from the
Stern-Gerlach probabilities
$p_m(\Omega)$ by replacing the weights in Eq.~\eqref{measurementDef} with 
the relevant parity operators $[M^R_0]_{mm}$ [see Fig.~\ref{RadonPlots}(a)-(b)]. One has
\begin{equation}
\label{RadonParityOperator}
M^R_{s} := \sum_{j=0}^{2J} \sqrt{\tfrac{2j+1}{4 \pi}} P_j(0)\allowbreak (\gamma_j)^{-s} \, \T_{j{}0}
\end{equation}
for general $s$-parametrized
phase-space representations 
where the Legendre polynomial $P_j(0)$ \cite{groemer1996} is used.
The point-symmetric parts $\tilde{W}_{|\mathrm{GHZ}\rangle}$ 
and $\tilde{W}_{|\mathrm{sq}\rangle}$ 
of the Wigner functions are 
recovered via an inverse spherical Radon transform
[right of Fig.~\ref{RadonPlots}(c)-(d)]. 
In general, one does however \emph{not} recover the complete Wigner function using this approach,
compare, e.g., the left and right part of Fig.~\ref{RadonPlots}(c).
But in typical experiments with large $J$, the Wigner function is localized around the north pole
and measuring probabilities $p_m(\Omega)$ close to the equator still allows for its
reconstruction from its Radon transform [middle of Fig.~\ref{RadonPlots}(d)]
by assuming a point-symmetric Wigner function [right of Fig.~\ref{RadonPlots}(d)],
which---in this particular case---still contains the full information of the quantum state,
cf.\ Sec.~3 in \cite{Schmied2011}.
The great-circle integrals for the spherical Radon transform
converge to the usual line integrals of the planar Radon transformation.

One concludes that the spherical Radon approach is \emph{not} suitable 
to recover general states of a finite-dimensional quantum system
due to essential geometric limitations of the spherical Radon transform.
This clarifies that not all infinite-dimensional
tomography schemes (as the Radon approach \cite{Smithey93,Leonhardt97,PR04,Brif98})
lead to unproblematic approaches when restricted to finite dimensions.

\subsection{Other aspects and discussion\label{tomo_discussion}}

As for infinite-dimensional phase-space methods (cf.\ Eq.~(6.8) in \cite{cahill1969}),
one can also use our approach to reconstruct the density matrix (which, however, is not the subject of this work)
\begin{equation}
\label{inversetrafo}
\rho =  \int_{S^2} F_\rho (\Omega,s)  \; \mathcal{R}(\Omega)  M_{-s}  \mathcal{R}^\dagger(\Omega)
\, \mathrm{d} \Omega,
\end{equation}
from its phase-space representation
by inverting Result~\ref{result1} with
a spherical integration.
Note that the reconstruction from the Q function is more precarious
as $M_{1}$ diverges for large $J$.
A tomography formula 
\begin{equation}\label{recovery_rho}
\rho = \sum_{m=-J}^J [M_s]_{mm}   \int_{S^2} p_m (\Omega)  \; 
\mathcal{R}(\Omega)  M_{-s}  \mathcal{R}^\dagger(\Omega)
\, \mathrm{d} \Omega
\end{equation}
in terms of the Stern-Gerlach probabilities $p_m (\Omega)$
is obtained by 
combining Eqs.~\eqref{measurementDef} and \eqref{inversetrafo},
where the integrals can be 
numerically estimated from finitely many spherical samples
via (e.g.) Gaussian quadratures \cite{kennedy2013book}. This generalizes 
\cite{amiet1998reconstructing,ariano2003spin,Manko97}, and the
`filtered backprojection' technique for the density matrix 
(see Eq.~(9) in \cite{Schmied2011}) agrees
in the limit of infinite measurements with Eq.~\eqref{recovery_rho}.

While a majority of earlier work focuses on reconstructing density matrices or infinite-dimensional phase-space functions
from measured data (see, e.g., \cite{schwemmer15,knips15,faist16,silva17,steffens17,riofrio17,suess16,PR04}),
we have presented in Eq.~\eqref{measurementDef} of
Result~\ref{result3} a general tomography formula for finite-dimensional
phase-space representations.
Such a tomography formula has not been reported before for the full class 
of all (finite-dimensional) $s$-parametrized phase-space representations.
Result~\ref{result3} provides the foundation
for engineering statistical estimators \cite{schervish} 
for the reconstruction of finite-dimensional
phase-space representations
in future research, which 
minimize the necessary Stern-Gerlach measurements
while guaranteeing robustness via precisely bounded confidence intervals
and ensuring a physical estimate. And Result~\ref{result4} provides a first step
in this direction.

For designing better statistical estimators,
the characterization of the type and relative size of
specific systematic and random errors involved 
in a given experimental realization would be necessary and choosing
a statistical estimator closely depends
on assumptions made in a concrete experiment.
Given a formula [as  Eq.~\eqref{measurementDef}]
to compute a desired target, one can use point estimators
(such as maximum likelihood estimators) or set estimators \cite{schervish} 
to determine a target which `best' fits to the measured data. 

The analysis in 
Sec.~\ref{subsec_sph_radon} shows that not all formulas used in the literature
produce the desired results, even before taking into account any  
statistical approach.
We have along these lines focussed in this work on the aspect of finding
suitable tomography formulas. Especially since related related experimental work has validated similar
tomography approaches relying on rotated parity operators (see Sec.~\ref{subsec_related})
and the statistical aspects are quite similar to the widely 
discussed cases of reconstructing density matrices or infinite-dimensional phase-space functions
\cite{schwemmer15,knips15,faist16,silva17,steffens17,riofrio17,suess16,PR04}.
In Secs.~\ref{sec:tomo}-\ref{simulations}, we have discussed 
the reconstruction errors that arise from having only 
a finite number $\nproj$ of Stern-Gerlach repetitions (i.e. shot noise).  
The resulting errors are illustrated in Fig.~\ref{reconstructions}
and they behave as expected.
The errors decrease as the number $\nproj$ of repetitions
increases. Beyond this first analysis, 
a more detailed discussion of statistical and robustness questions
is left to future work.
We want to only remark that reconstructing
a Wigner function directly using Result~\ref{result3} or Result~\ref{result4}
is---under noise---preferable to convolving/deconvolving noisy P or Q functions via Result~\ref{result2} as
convolutions are well known to be sensitive to noise
(cf.\ \cite{Agarwal98,Brif98}).
This claim is also substantiated by simulations of Stern-Gerlach tomographies
in Sec.~\ref{pitfalls} where the corresponding
reconstruction errors are also determined.
Therefore, not all reconstruction strategies 
are equally advisable under experimental noise
as detailed in Figs.~\ref{3x3histograms} and \ref{3x3histogramsl2}.
Concrete experiments will have to be explicitly designed  
depending on characteristics of the desired final
(phase-space) representation.

\section{Conclusion}

We have developed a unified formalism for
spherical phase-space
representations of finite-dimensional quantum states
based on rotated parity operators. 
The rotated parity operators appropriately reflect
the rotational symmetries of finite-dimensional quantum systems
and the Stern-Gerlach frequencies (or related overlaps)
from Eq.~\eqref{rotated_prob}
are easily measured in experiments (see Sec.~\ref{subsec_related}).
In addition, all of our results apply to the full class of 
(finite-dimensional) $s$-parametrized phase-space representations.
We have (a) systematically defined
spherical phase spaces for spin  systems
which recover 
the planar phase spaces from quantum optics 
in the large spin limit;
(b) different types of phase-space representations can be
translated into each other by convolving with spin-up
state representations;
(c) tomographic approaches 
can be now formulated consistently for all 
(finite-dimensional) $s$-parametrized phase-space representations;
(d) the spherical Radon approach is not suitable
to recover general states of finite-dimensional 
quantum systems.
Our results pave the way for innovative
tomography schemes 
to reconstruct phase-space functions 
of finite-dimensional quantum states.

\begin{acknowledgments}
The authors are especially grateful to
Thomas Schulte-Herbrüggen, Markus Grassl, Mark J.\ Everitt,
and Lukas Knips for discussions and comments.
B.K.\ acknowledges financial support
from the scholarship program of the 
Bavarian Academic Center for Central, Eastern and Southeastern Europe (BAYHOST)
and funding from the EU H2020-FETFLAG-03-2018 under grant agreement No 820495 (AQTION).
R.Z.\ and S.J.G.\ acknowledge support from the Deutsche Forschungsgemeinschaft  
through Grant No.\ Gl 203/7-2. This work is supported in part by the Elitenetzwerk Bayern through ExQM
and the Deutsche Forschungsgemeinschaft (DFG, German Research Foundation) 
under Germany’s Excellence Strategy -- EXC-2111 -- 39081486.
R.Z.\ acknowledges funding from the EU H2020-FETFLAG-03-2018 under grant agreement No 817482 (PASQuanS).
\end{acknowledgments}


\begin{thebibliography}{119}%
	\makeatletter
	\providecommand \@ifxundefined [1]{%
		\@ifx{#1\undefined}
	}%
	\providecommand \@ifnum [1]{%
		\ifnum #1\expandafter \@firstoftwo
		\else \expandafter \@secondoftwo
		\fi
	}%
	\providecommand \@ifx [1]{%
		\ifx #1\expandafter \@firstoftwo
		\else \expandafter \@secondoftwo
		\fi
	}%
	\providecommand \natexlab [1]{#1}%
	\providecommand \enquote  [1]{``#1''}%
	\providecommand \bibnamefont  [1]{#1}%
	\providecommand \bibfnamefont [1]{#1}%
	\providecommand \citenamefont [1]{#1}%
	\providecommand \href@noop [0]{\@secondoftwo}%
	\providecommand \href [0]{\begingroup \@sanitize@url \@href}%
	\providecommand \@href[1]{\@@startlink{#1}\@@href}%
	\providecommand \@@href[1]{\endgroup#1\@@endlink}%
	\providecommand \@sanitize@url [0]{\catcode `\\12\catcode `\$12\catcode
		`\&12\catcode `\#12\catcode `\^12\catcode `\_12\catcode `\%12\relax}%
	\providecommand \@@startlink[1]{}%
	\providecommand \@@endlink[0]{}%
	\providecommand \url  [0]{\begingroup\@sanitize@url \@url }%
	\providecommand \@url [1]{\endgroup\@href {#1}{\urlprefix }}%
	\providecommand \urlprefix  [0]{URL }%
	\providecommand \Eprint [0]{\href }%
	\providecommand \doibase [0]{http://dx.doi.org/}%
	\providecommand \selectlanguage [0]{\@gobble}%
	\providecommand \bibinfo  [0]{\@secondoftwo}%
	\providecommand \bibfield  [0]{\@secondoftwo}%
	\providecommand \translation [1]{[#1]}%
	\providecommand \BibitemOpen [0]{}%
	\providecommand \bibitemStop [0]{}%
	\providecommand \bibitemNoStop [0]{.\EOS\space}%
	\providecommand \EOS [0]{\spacefactor3000\relax}%
	\providecommand \BibitemShut  [1]{\csname bibitem#1\endcsname}%
	\let\auto@bib@innerbib\@empty
	\bibitem [{\citenamefont {Schleich}(2001)}]{SchleichBook}%
	\BibitemOpen
	\bibfield  {author} {\bibinfo {author} {\bibfnamefont {W.~P.}\ \bibnamefont
			{Schleich}},\ }\href@noop {} {\emph {\bibinfo {title} {{Quantum Optics in
					Phase Space}}}}\ (\bibinfo  {publisher} {Wiley-VCH, Berlin},\ \bibinfo {year}
	{2001})\BibitemShut {NoStop}%
	\bibitem [{\citenamefont {Zachos}\ \emph {et~al.}(2005)\citenamefont {Zachos},
		\citenamefont {Fairlie},\ and\ \citenamefont {Curtright}}]{zachos2005}%
	\BibitemOpen
	\bibfield  {author} {\bibinfo {author} {\bibfnamefont {C.~K.}\ \bibnamefont
			{Zachos}}, \bibinfo {author} {\bibfnamefont {D.~B.}\ \bibnamefont {Fairlie}},
		\ and\ \bibinfo {author} {\bibfnamefont {T.~L.}\ \bibnamefont {Curtright}},\
	}\href@noop {} {\emph {\bibinfo {title} {{Quantum Mechanics in Phase Space:
					An Overview with Selected Papers}}}}\ (\bibinfo  {publisher} {World
		Scientific, Singapore},\ \bibinfo {year} {2005})\BibitemShut {NoStop}%
	\bibitem [{\citenamefont {Schroeck~Jr.}(2013)}]{schroeck2013}%
	\BibitemOpen
	\bibfield  {author} {\bibinfo {author} {\bibfnamefont {F.~E.}\ \bibnamefont
			{Schroeck~Jr.}},\ }\href@noop {} {\emph {\bibinfo {title} {{Quantum Mechanics
					on Phase Space}}}}\ (\bibinfo  {publisher} {Springer, Dordrecht},\ \bibinfo
	{year} {2013})\BibitemShut {NoStop}%
	\bibitem [{\citenamefont {Curtright}\ \emph {et~al.}(2014)\citenamefont
		{Curtright}, \citenamefont {Fairlie},\ and\ \citenamefont
		{Zachos}}]{Curtright-review}%
	\BibitemOpen
	\bibfield  {author} {\bibinfo {author} {\bibfnamefont {T.~L.}\ \bibnamefont
			{Curtright}}, \bibinfo {author} {\bibfnamefont {D.~B.}\ \bibnamefont
			{Fairlie}}, \ and\ \bibinfo {author} {\bibfnamefont {C.~K.}\ \bibnamefont
			{Zachos}},\ }\href@noop {} {\emph {\bibinfo {title} {{A Concise Treatise on
					Quantum Mechanics in Phase Space}}}}\ (\bibinfo  {publisher} {World
		Scientific, Singapore},\ \bibinfo {year} {2014})\BibitemShut {NoStop}%
	\bibitem [{\citenamefont {Glauber}(1963)}]{Glauber1963}%
	\BibitemOpen
	\bibfield  {author} {\bibinfo {author} {\bibfnamefont {R.~J.}\ \bibnamefont
			{Glauber}},\ }\href {\doibase 10.1103/PhysRev.131.2766} {\bibfield  {journal}
		{\bibinfo  {journal} {Phys. Rev.}\ }\textbf {\bibinfo {volume} {131}},\
		\bibinfo {pages} {2766} (\bibinfo {year} {1963})}\BibitemShut {NoStop}%
	\bibitem [{\citenamefont {Glauber}(2006)}]{glauber2006nobel}%
	\BibitemOpen
	\bibfield  {author} {\bibinfo {author} {\bibfnamefont {R.~J.}\ \bibnamefont
			{Glauber}},\ }\href {\doibase 10.1103/RevModPhys.78.1267} {\bibfield
		{journal} {\bibinfo  {journal} {Rev. Mod. Phys.}\ }\textbf {\bibinfo {volume}
			{78}},\ \bibinfo {pages} {1267} (\bibinfo {year} {2006})}\BibitemShut
	{NoStop}%
	\bibitem [{\citenamefont {Cahill}\ and\ \citenamefont
		{Glauber}(1969)}]{cahill1969}%
	\BibitemOpen
	\bibfield  {author} {\bibinfo {author} {\bibfnamefont {K.~E.}\ \bibnamefont
			{Cahill}}\ and\ \bibinfo {author} {\bibfnamefont {R.~J.}\ \bibnamefont
			{Glauber}},\ }\href {\doibase 10.1103/PhysRev.177.1882} {\bibfield  {journal}
		{\bibinfo  {journal} {Phys. Rev.}\ }\textbf {\bibinfo {volume} {177}},\
		\bibinfo {pages} {1882} (\bibinfo {year} {1969})}\BibitemShut {NoStop}%
	\bibitem [{\citenamefont {Groenewold}(1946)}]{Gro46}%
	\BibitemOpen
	\bibfield  {author} {\bibinfo {author} {\bibfnamefont {H.}~\bibnamefont
			{Groenewold}},\ }\href {\doibase 10.1016/S0031-8914(46)80059-4} {\bibfield
		{journal} {\bibinfo  {journal} {Physica}\ }\textbf {\bibinfo {volume} {12}},\
		\bibinfo {pages} {405} (\bibinfo {year} {1946})}\BibitemShut {NoStop}%
	\bibitem [{\citenamefont {Moyal}(1949)}]{Moy49}%
	\BibitemOpen
	\bibfield  {author} {\bibinfo {author} {\bibfnamefont {J.~E.}\ \bibnamefont
			{Moyal}},\ }\href {\doibase 10.1017/S0305004100000487} {\bibfield  {journal}
		{\bibinfo  {journal} {Proc. Camb. Phil. Soc.}\ }\textbf {\bibinfo {volume}
			{45}},\ \bibinfo {pages} {99} (\bibinfo {year} {1949})}\BibitemShut {NoStop}%
	\bibitem [{\citenamefont {Bayen}\ \emph
		{et~al.}(1978{\natexlab{a}})\citenamefont {Bayen}, \citenamefont {Flato},
		\citenamefont {Fronsdal}, \citenamefont {Lichnerowicz},\ and\ \citenamefont
		{Sternheimer}}]{1bayen1978}%
	\BibitemOpen
	\bibfield  {author} {\bibinfo {author} {\bibfnamefont {F.}~\bibnamefont
			{Bayen}}, \bibinfo {author} {\bibfnamefont {M.}~\bibnamefont {Flato}},
		\bibinfo {author} {\bibfnamefont {C.}~\bibnamefont {Fronsdal}}, \bibinfo
		{author} {\bibfnamefont {A.}~\bibnamefont {Lichnerowicz}}, \ and\ \bibinfo
		{author} {\bibfnamefont {D.}~\bibnamefont {Sternheimer}},\ }\href {\doibase
		10.1016/0003-4916(78)90224-5} {\bibfield  {journal} {\bibinfo  {journal}
			{Ann. Phys.}\ }\textbf {\bibinfo {volume} {111}},\ \bibinfo {pages} {61}
		(\bibinfo {year} {1978}{\natexlab{a}})}\BibitemShut {NoStop}%
	\bibitem [{\citenamefont {Bayen}\ \emph
		{et~al.}(1978{\natexlab{b}})\citenamefont {Bayen}, \citenamefont {Flato},
		\citenamefont {Fronsdal}, \citenamefont {Lichnerowicz},\ and\ \citenamefont
		{Sternheimer}}]{2bayen1978}%
	\BibitemOpen
	\bibfield  {author} {\bibinfo {author} {\bibfnamefont {F.}~\bibnamefont
			{Bayen}}, \bibinfo {author} {\bibfnamefont {M.}~\bibnamefont {Flato}},
		\bibinfo {author} {\bibfnamefont {C.}~\bibnamefont {Fronsdal}}, \bibinfo
		{author} {\bibfnamefont {A.}~\bibnamefont {Lichnerowicz}}, \ and\ \bibinfo
		{author} {\bibfnamefont {D.}~\bibnamefont {Sternheimer}},\ }\href {\doibase
		10.1016/0003-4916(78)90225-7} {\bibfield  {journal} {\bibinfo  {journal}
			{Ann. Phys.}\ }\textbf {\bibinfo {volume} {111}},\ \bibinfo {pages} {111}
		(\bibinfo {year} {1978}{\natexlab{b}})}\BibitemShut {NoStop}%
	\bibitem [{\citenamefont {Berezin}(1974)}]{berezin74}%
	\BibitemOpen
	\bibfield  {author} {\bibinfo {author} {\bibfnamefont {F.~A.}\ \bibnamefont
			{Berezin}},\ }\href {\doibase 10.1070/IM1974v008n05ABEH002140} {\bibfield
		{journal} {\bibinfo  {journal} {Mathematics of the USSR-Izvestiya}\ }\textbf
		{\bibinfo {volume} {8}},\ \bibinfo {pages} {1109} (\bibinfo {year}
		{1974})}\BibitemShut {NoStop}%
	\bibitem [{\citenamefont {Berezin}(1975)}]{berezin75}%
	\BibitemOpen
	\bibfield  {author} {\bibinfo {author} {\bibfnamefont {F.~A.}\ \bibnamefont
			{Berezin}},\ }\href {\doibase 10.1007/BF01609397} {\bibfield  {journal}
		{\bibinfo  {journal} {Comm. Math. Phys.}\ }\textbf {\bibinfo {volume} {40}},\
		\bibinfo {pages} {153} (\bibinfo {year} {1975})}\BibitemShut {NoStop}%
	\bibitem [{\citenamefont {Weyl}(1927)}]{Wey27}%
	\BibitemOpen
	\bibfield  {author} {\bibinfo {author} {\bibfnamefont {H.}~\bibnamefont
			{Weyl}},\ }\href {\doibase 10.1007/BF02055756} {\bibfield  {journal}
		{\bibinfo  {journal} {Z. Phys.}\ }\textbf {\bibinfo {volume} {46}},\ \bibinfo
		{pages} {1} (\bibinfo {year} {1927})}\BibitemShut {NoStop}%
	\bibitem [{\citenamefont {Weyl}(1931)}]{Weyl31}%
	\BibitemOpen
	\bibfield  {author} {\bibinfo {author} {\bibfnamefont {H.}~\bibnamefont
			{Weyl}},\ }\href@noop {} {\emph {\bibinfo {title} {{Gruppentheorie und
					Quantenmechanik}}}},\ \bibinfo {edition} {2nd}\ ed.\ (\bibinfo  {publisher}
	{Hirzel, Leipzig},\ \bibinfo {year} {1931})\ \bibinfo {note} {english
		translation in \cite{Weyl50}}\BibitemShut {NoStop}%
	\bibitem [{\citenamefont {Weyl}(1950)}]{Weyl50}%
	\BibitemOpen
	\bibfield  {author} {\bibinfo {author} {\bibfnamefont {H.}~\bibnamefont
			{Weyl}},\ }\href@noop {} {\emph {\bibinfo {title} {{The Theory of Groups \&
					Quantum Mechanics}}}},\ \bibinfo {edition} {2nd}\ ed.\ (\bibinfo  {publisher}
	{Dover Publ., New York},\ \bibinfo {year} {1950})\BibitemShut {NoStop}%
	\bibitem [{\citenamefont {de~Gosson}(2017)}]{thewignertransform}%
	\BibitemOpen
	\bibfield  {author} {\bibinfo {author} {\bibfnamefont {M.}~\bibnamefont
			{de~Gosson}},\ }\href@noop {} {\emph {\bibinfo {title} {{The Wigner
					Transform}}}}\ (\bibinfo  {publisher} {World Scientific, London},\ \bibinfo
	{year} {2017})\BibitemShut {NoStop}%
	\bibitem [{\citenamefont {de~Gosson}(2016)}]{deGosson2016}%
	\BibitemOpen
	\bibfield  {author} {\bibinfo {author} {\bibfnamefont {M.~A.}\ \bibnamefont
			{de~Gosson}},\ }\enquote {\bibinfo {title} {{Born-Jordan Quantization: Theory
				and Applications}},}\ \ (\bibinfo  {publisher} {Springer, Switzerland},\
	\bibinfo {year} {2016})\ pp.\ \bibinfo {pages} {113--127}\BibitemShut
	{NoStop}%
	\bibitem [{\citenamefont {Gr{\"o}chenig}(2001)}]{groechenig2001foundations}%
	\BibitemOpen
	\bibfield  {author} {\bibinfo {author} {\bibfnamefont {K.}~\bibnamefont
			{Gr{\"o}chenig}},\ }\href@noop {} {\emph {\bibinfo {title} {Foundations of
				Time-Frequency Analysis}}}\ (\bibinfo  {publisher} {Birkh{\"a}user, Boston},\
	\bibinfo {year} {2001})\BibitemShut {NoStop}%
	\bibitem [{\citenamefont {Cohen}(1966)}]{cohen1966generalized}%
	\BibitemOpen
	\bibfield  {author} {\bibinfo {author} {\bibfnamefont {L.}~\bibnamefont
			{Cohen}},\ }\href {\doibase 10.1063/1.1931206} {\bibfield  {journal}
		{\bibinfo  {journal} {J. Math. Phys.}\ }\textbf {\bibinfo {volume} {7}},\
		\bibinfo {pages} {781} (\bibinfo {year} {1966})}\BibitemShut {NoStop}%
	\bibitem [{\citenamefont {Cohen}(1995)}]{Cohen95}%
	\BibitemOpen
	\bibfield  {author} {\bibinfo {author} {\bibfnamefont {L.}~\bibnamefont
			{Cohen}},\ }\href@noop {} {\emph {\bibinfo {title} {{Time-Frequency
					Analysis}}}}\ (\bibinfo  {publisher} {Prentice-Hall, Englewood Cliffs, NJ},\
	\bibinfo {year} {1995})\BibitemShut {NoStop}%
	\bibitem [{\citenamefont {McConnell}\ \emph {et~al.}(2015)\citenamefont
		{McConnell}, \citenamefont {Zhang}, \citenamefont {Hu}, \citenamefont
		{{\'C}uk},\ and\ \citenamefont {Vuleti{\'c}}}]{mcconnell2015}%
	\BibitemOpen
	\bibfield  {author} {\bibinfo {author} {\bibfnamefont {R.}~\bibnamefont
			{McConnell}}, \bibinfo {author} {\bibfnamefont {H.}~\bibnamefont {Zhang}},
		\bibinfo {author} {\bibfnamefont {J.}~\bibnamefont {Hu}}, \bibinfo {author}
		{\bibfnamefont {S.}~\bibnamefont {{\'C}uk}}, \ and\ \bibinfo {author}
		{\bibfnamefont {V.}~\bibnamefont {Vuleti{\'c}}},\ }\href {\doibase
		10.1038/nature14293} {\bibfield  {journal} {\bibinfo  {journal} {Nature}\
		}\textbf {\bibinfo {volume} {519}},\ \bibinfo {pages} {439} (\bibinfo {year}
		{2015})}\BibitemShut {NoStop}%
	\bibitem [{\citenamefont {Haas}\ \emph {et~al.}(2014)\citenamefont {Haas},
		\citenamefont {Volz}, \citenamefont {Gehr}, \citenamefont {Reichel},\ and\
		\citenamefont {Est{\`e}ve}}]{haas2014}%
	\BibitemOpen
	\bibfield  {author} {\bibinfo {author} {\bibfnamefont {F.}~\bibnamefont
			{Haas}}, \bibinfo {author} {\bibfnamefont {J.}~\bibnamefont {Volz}}, \bibinfo
		{author} {\bibfnamefont {R.}~\bibnamefont {Gehr}}, \bibinfo {author}
		{\bibfnamefont {J.}~\bibnamefont {Reichel}}, \ and\ \bibinfo {author}
		{\bibfnamefont {J.}~\bibnamefont {Est{\`e}ve}},\ }\href {\doibase
		10.1126/science.1248905} {\bibfield  {journal} {\bibinfo  {journal}
			{Science}\ }\textbf {\bibinfo {volume} {344}},\ \bibinfo {pages} {180}
		(\bibinfo {year} {2014})}\BibitemShut {NoStop}%
	\bibitem [{\citenamefont {Anderson}\ \emph {et~al.}(1995)\citenamefont
		{Anderson}, \citenamefont {Ensher}, \citenamefont {Matthews}, \citenamefont
		{Wieman},\ and\ \citenamefont {Cornell}}]{anderson1995}%
	\BibitemOpen
	\bibfield  {author} {\bibinfo {author} {\bibfnamefont {M.~H.}\ \bibnamefont
			{Anderson}}, \bibinfo {author} {\bibfnamefont {J.~R.}\ \bibnamefont
			{Ensher}}, \bibinfo {author} {\bibfnamefont {M.~R.}\ \bibnamefont
			{Matthews}}, \bibinfo {author} {\bibfnamefont {C.~E.}\ \bibnamefont
			{Wieman}}, \ and\ \bibinfo {author} {\bibfnamefont {E.~A.}\ \bibnamefont
			{Cornell}},\ }\href {\doibase 10.1126/science.269.5221.198} {\bibfield
		{journal} {\bibinfo  {journal} {Science}\ }\textbf {\bibinfo {volume}
			{269}},\ \bibinfo {pages} {198} (\bibinfo {year} {1995})}\BibitemShut
	{NoStop}%
	\bibitem [{\citenamefont {Ho}(1998)}]{ho1998}%
	\BibitemOpen
	\bibfield  {author} {\bibinfo {author} {\bibfnamefont {T.-L.}\ \bibnamefont
			{Ho}},\ }\href {\doibase 10.1103/PhysRevLett.81.742} {\bibfield  {journal}
		{\bibinfo  {journal} {Phys. Rev. Lett.}\ }\textbf {\bibinfo {volume} {81}},\
		\bibinfo {pages} {742} (\bibinfo {year} {1998})}\BibitemShut {NoStop}%
	\bibitem [{\citenamefont {Ohmi}\ and\ \citenamefont
		{Machida}(1998)}]{ohmi1998}%
	\BibitemOpen
	\bibfield  {author} {\bibinfo {author} {\bibfnamefont {T.}~\bibnamefont
			{Ohmi}}\ and\ \bibinfo {author} {\bibfnamefont {K.}~\bibnamefont {Machida}},\
	}\href {\doibase 10.1143/JPSJ.67.1822} {\bibfield  {journal} {\bibinfo
			{journal} {J. Phys. Soc. Jpn.}\ }\textbf {\bibinfo {volume} {67}},\ \bibinfo
		{pages} {1822} (\bibinfo {year} {1998})}\BibitemShut {NoStop}%
	\bibitem [{\citenamefont {Stenger}\ \emph {et~al.}(1998)\citenamefont
		{Stenger}, \citenamefont {Inouye}, \citenamefont {Stamper-Kurn},
		\citenamefont {Miesner}, \citenamefont {Chikkatur},\ and\ \citenamefont
		{Ketterle}}]{stenger1999}%
	\BibitemOpen
	\bibfield  {author} {\bibinfo {author} {\bibfnamefont {J.}~\bibnamefont
			{Stenger}}, \bibinfo {author} {\bibfnamefont {S.}~\bibnamefont {Inouye}},
		\bibinfo {author} {\bibfnamefont {D.}~\bibnamefont {Stamper-Kurn}}, \bibinfo
		{author} {\bibfnamefont {H.-J.}\ \bibnamefont {Miesner}}, \bibinfo {author}
		{\bibfnamefont {A.}~\bibnamefont {Chikkatur}}, \ and\ \bibinfo {author}
		{\bibfnamefont {W.}~\bibnamefont {Ketterle}},\ }\href {\doibase
		10.1038/24567} {\bibfield  {journal} {\bibinfo  {journal} {Nature}\ }\textbf
		{\bibinfo {volume} {396}},\ \bibinfo {pages} {345} (\bibinfo {year}
		{1998})}\BibitemShut {NoStop}%
	\bibitem [{\citenamefont {Lin}\ \emph {et~al.}(2011)\citenamefont {Lin},
		\citenamefont {Jim{\'e}nez-Garc{\'\i}a},\ and\ \citenamefont
		{Spielman}}]{lin2011}%
	\BibitemOpen
	\bibfield  {author} {\bibinfo {author} {\bibfnamefont {Y.-J.}\ \bibnamefont
			{Lin}}, \bibinfo {author} {\bibfnamefont {K.}~\bibnamefont
			{Jim{\'e}nez-Garc{\'\i}a}}, \ and\ \bibinfo {author} {\bibfnamefont
			{I.}~\bibnamefont {Spielman}},\ }\href {\doibase 10.1038/nature09887}
	{\bibfield  {journal} {\bibinfo  {journal} {Nature}\ }\textbf {\bibinfo
			{volume} {471}},\ \bibinfo {pages} {83} (\bibinfo {year} {2011})}\BibitemShut
	{NoStop}%
	\bibitem [{\citenamefont {Riedel}\ \emph {et~al.}(2010)\citenamefont {Riedel},
		\citenamefont {Böhi}, \citenamefont {Li}, \citenamefont {Hänsch},
		\citenamefont {Sinatra},\ and\ \citenamefont {Treutlein}}]{treutlein2010}%
	\BibitemOpen
	\bibfield  {author} {\bibinfo {author} {\bibfnamefont {M.~F.}\ \bibnamefont
			{Riedel}}, \bibinfo {author} {\bibfnamefont {P.}~\bibnamefont {Böhi}},
		\bibinfo {author} {\bibfnamefont {Y.}~\bibnamefont {Li}}, \bibinfo {author}
		{\bibfnamefont {T.~W.}\ \bibnamefont {Hänsch}}, \bibinfo {author}
		{\bibfnamefont {A.}~\bibnamefont {Sinatra}}, \ and\ \bibinfo {author}
		{\bibfnamefont {P.}~\bibnamefont {Treutlein}},\ }\href {\doibase
		10.1038/nature08988} {\bibfield  {journal} {\bibinfo  {journal} {Nature}\
		}\textbf {\bibinfo {volume} {464}},\ \bibinfo {pages} {1170} (\bibinfo {year}
		{2010})}\BibitemShut {NoStop}%
	\bibitem [{\citenamefont {Schmied}\ and\ \citenamefont
		{Treutlein}(2011)}]{Schmied2011}%
	\BibitemOpen
	\bibfield  {author} {\bibinfo {author} {\bibfnamefont {R.}~\bibnamefont
			{Schmied}}\ and\ \bibinfo {author} {\bibfnamefont {P.}~\bibnamefont
			{Treutlein}},\ }\href {\doibase 10.1088/1367-2630/13/6/065019} {\bibfield
		{journal} {\bibinfo  {journal} {New J. Phys.}\ }\textbf {\bibinfo {volume}
			{13}},\ \bibinfo {pages} {065019} (\bibinfo {year} {2011})}\BibitemShut
	{NoStop}%
	\bibitem [{\citenamefont {Hamley}\ \emph {et~al.}(2012)\citenamefont {Hamley},
		\citenamefont {Gerving}, \citenamefont {Hoang}, \citenamefont {Bookjans},\
		and\ \citenamefont {Chapman}}]{hamley2012}%
	\BibitemOpen
	\bibfield  {author} {\bibinfo {author} {\bibfnamefont {C.~D.}\ \bibnamefont
			{Hamley}}, \bibinfo {author} {\bibfnamefont {C.~S.}\ \bibnamefont {Gerving}},
		\bibinfo {author} {\bibfnamefont {T.~M.}\ \bibnamefont {Hoang}}, \bibinfo
		{author} {\bibfnamefont {E.~M.}\ \bibnamefont {Bookjans}}, \ and\ \bibinfo
		{author} {\bibfnamefont {M.~S.}\ \bibnamefont {Chapman}},\ }\href {\doibase
		10.1038/nphys2245} {\bibfield  {journal} {\bibinfo  {journal} {Nat. Phys.}\
		}\textbf {\bibinfo {volume} {8}},\ \bibinfo {pages} {305} (\bibinfo {year}
		{2012})}\BibitemShut {NoStop}%
	\bibitem [{\citenamefont {Strobel}\ \emph {et~al.}(2014)\citenamefont
		{Strobel}, \citenamefont {Muessel}, \citenamefont {Linnemann}, \citenamefont
		{Zibold}, \citenamefont {Hume}, \citenamefont {Pezz{\`e}}, \citenamefont
		{Smerzi},\ and\ \citenamefont {Oberthaler}}]{strobel2014}%
	\BibitemOpen
	\bibfield  {author} {\bibinfo {author} {\bibfnamefont {H.}~\bibnamefont
			{Strobel}}, \bibinfo {author} {\bibfnamefont {W.}~\bibnamefont {Muessel}},
		\bibinfo {author} {\bibfnamefont {D.}~\bibnamefont {Linnemann}}, \bibinfo
		{author} {\bibfnamefont {T.}~\bibnamefont {Zibold}}, \bibinfo {author}
		{\bibfnamefont {D.~B.}\ \bibnamefont {Hume}}, \bibinfo {author}
		{\bibfnamefont {L.}~\bibnamefont {Pezz{\`e}}}, \bibinfo {author}
		{\bibfnamefont {A.}~\bibnamefont {Smerzi}}, \ and\ \bibinfo {author}
		{\bibfnamefont {M.~K.}\ \bibnamefont {Oberthaler}},\ }\href {\doibase
		10.1126/science.1250147} {\bibfield  {journal} {\bibinfo  {journal}
			{Science}\ }\textbf {\bibinfo {volume} {345}},\ \bibinfo {pages} {424}
		(\bibinfo {year} {2014})}\BibitemShut {NoStop}%
	\bibitem [{\citenamefont {Leibfried}\ \emph {et~al.}(2005)\citenamefont
		{Leibfried}, \citenamefont {Knill}, \citenamefont {Seidelin}, \citenamefont
		{Britton}, \citenamefont {Blakestad}, \citenamefont {Chiaverini},
		\citenamefont {Hume}, \citenamefont {Itano}, \citenamefont {Jost},
		\citenamefont {Langer}, \citenamefont {Reichle},\ and\ \citenamefont
		{Wineland}}]{leibfried2005}%
	\BibitemOpen
	\bibfield  {author} {\bibinfo {author} {\bibfnamefont {D.}~\bibnamefont
			{Leibfried}}, \bibinfo {author} {\bibfnamefont {E.}~\bibnamefont {Knill}},
		\bibinfo {author} {\bibfnamefont {S.}~\bibnamefont {Seidelin}}, \bibinfo
		{author} {\bibfnamefont {J.}~\bibnamefont {Britton}}, \bibinfo {author}
		{\bibfnamefont {R.~B.}\ \bibnamefont {Blakestad}}, \bibinfo {author}
		{\bibfnamefont {J.}~\bibnamefont {Chiaverini}}, \bibinfo {author}
		{\bibfnamefont {D.~B.}\ \bibnamefont {Hume}}, \bibinfo {author}
		{\bibfnamefont {W.~M.}\ \bibnamefont {Itano}}, \bibinfo {author}
		{\bibfnamefont {J.~D.}\ \bibnamefont {Jost}}, \bibinfo {author}
		{\bibfnamefont {C.}~\bibnamefont {Langer}}, \bibinfo {author} {\bibfnamefont
			{R.}~\bibnamefont {Reichle}}, \ and\ \bibinfo {author} {\bibfnamefont
			{D.~J.}\ \bibnamefont {Wineland}},\ }\href {\doibase 10.1038/nature04251}
	{\bibfield  {journal} {\bibinfo  {journal} {Nature}\ }\textbf {\bibinfo
			{volume} {438}},\ \bibinfo {pages} {639} (\bibinfo {year}
		{2005})}\BibitemShut {NoStop}%
	\bibitem [{\citenamefont {Bohnet}\ \emph {et~al.}(2016)\citenamefont {Bohnet},
		\citenamefont {Sawyer}, \citenamefont {Britton}, \citenamefont {Wall},
		\citenamefont {Rey}, \citenamefont {Foss-Feig},\ and\ \citenamefont
		{Bollinger}}]{bohnet2016}%
	\BibitemOpen
	\bibfield  {author} {\bibinfo {author} {\bibfnamefont {J.~G.}\ \bibnamefont
			{Bohnet}}, \bibinfo {author} {\bibfnamefont {B.~C.}\ \bibnamefont {Sawyer}},
		\bibinfo {author} {\bibfnamefont {J.~W.}\ \bibnamefont {Britton}}, \bibinfo
		{author} {\bibfnamefont {M.~L.}\ \bibnamefont {Wall}}, \bibinfo {author}
		{\bibfnamefont {A.~M.}\ \bibnamefont {Rey}}, \bibinfo {author} {\bibfnamefont
			{M.}~\bibnamefont {Foss-Feig}}, \ and\ \bibinfo {author} {\bibfnamefont
			{J.~J.}\ \bibnamefont {Bollinger}},\ }\href {\doibase
		10.1126/science.aad9958} {\bibfield  {journal} {\bibinfo  {journal}
			{Science}\ }\textbf {\bibinfo {volume} {352}},\ \bibinfo {pages} {1297}
		(\bibinfo {year} {2016})}\BibitemShut {NoStop}%
	\bibitem [{\citenamefont {Monz}\ \emph {et~al.}(2011)\citenamefont {Monz},
		\citenamefont {Schindler}, \citenamefont {Barreiro}, \citenamefont {Chwalla},
		\citenamefont {Nigg}, \citenamefont {Coish}, \citenamefont {Harlander},
		\citenamefont {H{\"a}nsel}, \citenamefont {Hennrich},\ and\ \citenamefont
		{Blatt}}]{monz2011}%
	\BibitemOpen
	\bibfield  {author} {\bibinfo {author} {\bibfnamefont {T.}~\bibnamefont
			{Monz}}, \bibinfo {author} {\bibfnamefont {P.}~\bibnamefont {Schindler}},
		\bibinfo {author} {\bibfnamefont {J.~T.}\ \bibnamefont {Barreiro}}, \bibinfo
		{author} {\bibfnamefont {M.}~\bibnamefont {Chwalla}}, \bibinfo {author}
		{\bibfnamefont {D.}~\bibnamefont {Nigg}}, \bibinfo {author} {\bibfnamefont
			{W.~A.}\ \bibnamefont {Coish}}, \bibinfo {author} {\bibfnamefont
			{M.}~\bibnamefont {Harlander}}, \bibinfo {author} {\bibfnamefont
			{W.}~\bibnamefont {H{\"a}nsel}}, \bibinfo {author} {\bibfnamefont
			{M.}~\bibnamefont {Hennrich}}, \ and\ \bibinfo {author} {\bibfnamefont
			{R.}~\bibnamefont {Blatt}},\ }\href {\doibase 10.1103/PhysRevLett.106.130506}
	{\bibfield  {journal} {\bibinfo  {journal} {Phys. Rev. Lett.}\ }\textbf
		{\bibinfo {volume} {106}},\ \bibinfo {pages} {130506} (\bibinfo {year}
		{2011})}\BibitemShut {NoStop}%
	\bibitem [{\citenamefont {Bouchard}\ \emph {et~al.}(2017)\citenamefont
		{Bouchard}, \citenamefont {de~la Hoz}, \citenamefont {Bjork}, \citenamefont
		{Boyd}, \citenamefont {Grassl}, \citenamefont {Hradil}, \citenamefont
		{Karimi}, \citenamefont {Klimov}, \citenamefont {Leuchs}, \citenamefont
		{Rehacek},\ and\ \citenamefont {Sanchez-Soto}}]{bouchard2016}%
	\BibitemOpen
	\bibfield  {author} {\bibinfo {author} {\bibfnamefont {F.}~\bibnamefont
			{Bouchard}}, \bibinfo {author} {\bibfnamefont {P.}~\bibnamefont {de~la Hoz}},
		\bibinfo {author} {\bibfnamefont {G.}~\bibnamefont {Bjork}}, \bibinfo
		{author} {\bibfnamefont {R.~W.}\ \bibnamefont {Boyd}}, \bibinfo {author}
		{\bibfnamefont {M.}~\bibnamefont {Grassl}}, \bibinfo {author} {\bibfnamefont
			{Z.}~\bibnamefont {Hradil}}, \bibinfo {author} {\bibfnamefont
			{E.}~\bibnamefont {Karimi}}, \bibinfo {author} {\bibfnamefont
			{A.}~\bibnamefont {Klimov}}, \bibinfo {author} {\bibfnamefont
			{G.}~\bibnamefont {Leuchs}}, \bibinfo {author} {\bibfnamefont
			{J.}~\bibnamefont {Rehacek}}, \ and\ \bibinfo {author} {\bibfnamefont
			{L.~L.}\ \bibnamefont {Sanchez-Soto}},\ }\href {\doibase
		10.1364/OPTICA.4.001429} {\bibfield  {journal} {\bibinfo  {journal} {Optica}\
		}\textbf {\bibinfo {volume} {4}},\ \bibinfo {pages} {1429} (\bibinfo {year}
		{2017})}\BibitemShut {NoStop}%
	\bibitem [{\citenamefont {Klimov}\ \emph
		{et~al.}(2017{\natexlab{a}})\citenamefont {Klimov}, \citenamefont {Zwierz},
		\citenamefont {Wallentowitz}, \citenamefont {Jarzyna},\ and\ \citenamefont
		{Banaszek}}]{klimov2017}%
	\BibitemOpen
	\bibfield  {author} {\bibinfo {author} {\bibfnamefont {A.~B.}\ \bibnamefont
			{Klimov}}, \bibinfo {author} {\bibfnamefont {M.}~\bibnamefont {Zwierz}},
		\bibinfo {author} {\bibfnamefont {S.}~\bibnamefont {Wallentowitz}}, \bibinfo
		{author} {\bibfnamefont {M.}~\bibnamefont {Jarzyna}}, \ and\ \bibinfo
		{author} {\bibfnamefont {K.}~\bibnamefont {Banaszek}},\ }\href {\doibase
		110.1088/1367-2630/aa73b0} {\bibfield  {journal} {\bibinfo  {journal} {New J.
				Phys}\ }\textbf {\bibinfo {volume} {19}},\ \bibinfo {pages} {073013}
		(\bibinfo {year} {2017}{\natexlab{a}})}\BibitemShut {NoStop}%
	\bibitem [{\citenamefont {Chaturvedi}\ \emph {et~al.}(2006)\citenamefont
		{Chaturvedi}, \citenamefont {Marmo}, \citenamefont {Mukunda}, \citenamefont
		{Simon},\ and\ \citenamefont {Zampini}}]{chaturvedi2006}%
	\BibitemOpen
	\bibfield  {author} {\bibinfo {author} {\bibfnamefont {S.}~\bibnamefont
			{Chaturvedi}}, \bibinfo {author} {\bibfnamefont {G.}~\bibnamefont {Marmo}},
		\bibinfo {author} {\bibfnamefont {N.}~\bibnamefont {Mukunda}}, \bibinfo
		{author} {\bibfnamefont {R.}~\bibnamefont {Simon}}, \ and\ \bibinfo {author}
		{\bibfnamefont {A.}~\bibnamefont {Zampini}},\ }\href {\doibase
		10.1142/S0129055X06002802} {\bibfield  {journal} {\bibinfo  {journal} {Rev.
				Math. Phys.}\ }\textbf {\bibinfo {volume} {18}},\ \bibinfo {pages} {887}
		(\bibinfo {year} {2006})}\BibitemShut {NoStop}%
	\bibitem [{\citenamefont {Wootters}(1987)}]{Wootters87}%
	\BibitemOpen
	\bibfield  {author} {\bibinfo {author} {\bibfnamefont {W.~K.}\ \bibnamefont
			{Wootters}},\ }\href {\doibase 10.1016/0003-4916(87)90176-X} {\bibfield
		{journal} {\bibinfo  {journal} {Ann. Phys.}\ }\textbf {\bibinfo {volume}
			{176}},\ \bibinfo {pages} {1} (\bibinfo {year} {1987})}\BibitemShut {NoStop}%
	\bibitem [{\citenamefont {Leonhardt}(1996)}]{leonhardt1996}%
	\BibitemOpen
	\bibfield  {author} {\bibinfo {author} {\bibfnamefont {U.}~\bibnamefont
			{Leonhardt}},\ }\href {\doibase 10.1103/PhysRevA.53.2998} {\bibfield
		{journal} {\bibinfo  {journal} {Phys. Rev. A}\ }\textbf {\bibinfo {volume}
			{53}},\ \bibinfo {pages} {2998} (\bibinfo {year} {1996})}\BibitemShut
	{NoStop}%
	\bibitem [{\citenamefont {Gibbons}\ \emph {et~al.}(2004)\citenamefont
		{Gibbons}, \citenamefont {Hoffman},\ and\ \citenamefont
		{Wootters}}]{gibbons2004}%
	\BibitemOpen
	\bibfield  {author} {\bibinfo {author} {\bibfnamefont {K.~S.}\ \bibnamefont
			{Gibbons}}, \bibinfo {author} {\bibfnamefont {M.~J.}\ \bibnamefont
			{Hoffman}}, \ and\ \bibinfo {author} {\bibfnamefont {W.~K.}\ \bibnamefont
			{Wootters}},\ }\href {\doibase 10.1103/PhysRevA.70.062101} {\bibfield
		{journal} {\bibinfo  {journal} {Phys. Rev. A}\ }\textbf {\bibinfo {volume}
			{70}},\ \bibinfo {pages} {062101} (\bibinfo {year} {2004})}\BibitemShut
	{NoStop}%
	\bibitem [{\citenamefont {Ferrie}\ and\ \citenamefont
		{Emerson}(2009)}]{ferrie2009}%
	\BibitemOpen
	\bibfield  {author} {\bibinfo {author} {\bibfnamefont {C.}~\bibnamefont
			{Ferrie}}\ and\ \bibinfo {author} {\bibfnamefont {J.}~\bibnamefont
			{Emerson}},\ }\href {\doibase 10.1088/1367-2630/11/6/063040} {\bibfield
		{journal} {\bibinfo  {journal} {New J. Phys.}\ }\textbf {\bibinfo {volume}
			{11}},\ \bibinfo {pages} {063040} (\bibinfo {year} {2009})}\BibitemShut
	{NoStop}%
	\bibitem [{\citenamefont {Leonhardt}(1997)}]{Leonhardt97}%
	\BibitemOpen
	\bibfield  {author} {\bibinfo {author} {\bibfnamefont {U.}~\bibnamefont
			{Leonhardt}},\ }\href@noop {} {\emph {\bibinfo {title} {{Measuring the
					Quantum State of Light}}}}\ (\bibinfo  {publisher} {Cambridge Univ.\ Press,
		Cambridge},\ \bibinfo {year} {1997})\BibitemShut {NoStop}%
	\bibitem [{\citenamefont {Smithey}\ \emph {et~al.}(1993)\citenamefont
		{Smithey}, \citenamefont {Beck}, \citenamefont {Raymer},\ and\ \citenamefont
		{Faridani}}]{Smithey93}%
	\BibitemOpen
	\bibfield  {author} {\bibinfo {author} {\bibfnamefont {D.~T.}\ \bibnamefont
			{Smithey}}, \bibinfo {author} {\bibfnamefont {M.}~\bibnamefont {Beck}},
		\bibinfo {author} {\bibfnamefont {M.~G.}\ \bibnamefont {Raymer}}, \ and\
		\bibinfo {author} {\bibfnamefont {A.}~\bibnamefont {Faridani}},\ }\href
	{\doibase 10.1103/PhysRevLett.70.1244} {\bibfield  {journal} {\bibinfo
			{journal} {Phys. Rev. Lett.}\ }\textbf {\bibinfo {volume} {70}},\ \bibinfo
		{pages} {1244} (\bibinfo {year} {1993})}\BibitemShut {NoStop}%
	\bibitem [{\citenamefont {Husimi}(1940)}]{husimi1940}%
	\BibitemOpen
	\bibfield  {author} {\bibinfo {author} {\bibfnamefont {K.}~\bibnamefont
			{Husimi}},\ }\href {\doibase 10.11429/ppmsj1919.22.4_264} {\bibfield
		{journal} {\bibinfo  {journal} {Proc. Phys. Math. Soc. Japan}\ }\textbf
		{\bibinfo {volume} {22}},\ \bibinfo {pages} {264} (\bibinfo {year}
		{1940})}\BibitemShut {NoStop}%
	\bibitem [{\citenamefont {Kanem}\ \emph {et~al.}(2005)\citenamefont {Kanem},
		\citenamefont {Maneshi}, \citenamefont {Myrskog},\ and\ \citenamefont
		{Steinberg}}]{kanem2005}%
	\BibitemOpen
	\bibfield  {author} {\bibinfo {author} {\bibfnamefont {J.}~\bibnamefont
			{Kanem}}, \bibinfo {author} {\bibfnamefont {S.}~\bibnamefont {Maneshi}},
		\bibinfo {author} {\bibfnamefont {S.}~\bibnamefont {Myrskog}}, \ and\
		\bibinfo {author} {\bibfnamefont {A.}~\bibnamefont {Steinberg}},\ }\href
	{\doibase 10.1088/1464-4266/7/12/037} {\bibfield  {journal} {\bibinfo
			{journal} {J. Opt. B}\ }\textbf {\bibinfo {volume} {7}},\ \bibinfo {pages}
		{S705} (\bibinfo {year} {2005})}\BibitemShut {NoStop}%
	\bibitem [{\citenamefont {Eichler}\ \emph {et~al.}(2011)\citenamefont
		{Eichler}, \citenamefont {Bozyigit}, \citenamefont {Lang}, \citenamefont
		{Steffen}, \citenamefont {Fink},\ and\ \citenamefont {Wallraff}}]{Eichler11}%
	\BibitemOpen
	\bibfield  {author} {\bibinfo {author} {\bibfnamefont {C.}~\bibnamefont
			{Eichler}}, \bibinfo {author} {\bibfnamefont {D.}~\bibnamefont {Bozyigit}},
		\bibinfo {author} {\bibfnamefont {C.}~\bibnamefont {Lang}}, \bibinfo {author}
		{\bibfnamefont {L.}~\bibnamefont {Steffen}}, \bibinfo {author} {\bibfnamefont
			{J.}~\bibnamefont {Fink}}, \ and\ \bibinfo {author} {\bibfnamefont
			{A.}~\bibnamefont {Wallraff}},\ }\href {\doibase
		10.1103/PhysRevLett.106.220503} {\bibfield  {journal} {\bibinfo  {journal}
			{Phys. Rev. Lett.}\ }\textbf {\bibinfo {volume} {106}},\ \bibinfo {pages}
		{220503} (\bibinfo {year} {2011})}\BibitemShut {NoStop}%
	\bibitem [{\citenamefont {Agarwal}(1998)}]{Agarwal98}%
	\BibitemOpen
	\bibfield  {author} {\bibinfo {author} {\bibfnamefont {G.~S.}\ \bibnamefont
			{Agarwal}},\ }\href {\doibase 10.1103/PhysRevA.57.671} {\bibfield  {journal}
		{\bibinfo  {journal} {Phys. Rev. A}\ }\textbf {\bibinfo {volume} {57}},\
		\bibinfo {pages} {671} (\bibinfo {year} {1998})}\BibitemShut {NoStop}%
	\bibitem [{\citenamefont {Koczor}\ \emph
		{et~al.}(2019{\natexlab{a}})\citenamefont {Koczor}, \citenamefont {Zeier},\
		and\ \citenamefont {Glaser}}]{KZG}%
	\BibitemOpen
	\bibfield  {author} {\bibinfo {author} {\bibfnamefont {B.}~\bibnamefont
			{Koczor}}, \bibinfo {author} {\bibfnamefont {R.}~\bibnamefont {Zeier}}, \
		and\ \bibinfo {author} {\bibfnamefont {S.~J.}\ \bibnamefont {Glaser}},\
	}\href {\doibase 10.1088/1751-8121/aaf302} {\bibfield  {journal} {\bibinfo
			{journal} {J. Phys. A.}\ }\textbf {\bibinfo {volume} {52}},\ \bibinfo {pages}
		{055302} (\bibinfo {year} {2019}{\natexlab{a}})}\BibitemShut {NoStop}%
	\bibitem [{\citenamefont {Dicke}(1954)}]{Dicke1954}%
	\BibitemOpen
	\bibfield  {author} {\bibinfo {author} {\bibfnamefont {R.~H.}\ \bibnamefont
			{Dicke}},\ }\href {\doibase 10.1103/PhysRev.93.99} {\bibfield  {journal}
		{\bibinfo  {journal} {Phys. Rev.}\ }\textbf {\bibinfo {volume} {93}},\
		\bibinfo {pages} {99} (\bibinfo {year} {1954})}\BibitemShut {NoStop}%
	\bibitem [{\citenamefont {Stockton}\ \emph {et~al.}(2003)\citenamefont
		{Stockton}, \citenamefont {Geremia}, \citenamefont {Doherty},\ and\
		\citenamefont {Mabuchi}}]{stockton2003}%
	\BibitemOpen
	\bibfield  {author} {\bibinfo {author} {\bibfnamefont {J.~K.}\ \bibnamefont
			{Stockton}}, \bibinfo {author} {\bibfnamefont {J.~M.}\ \bibnamefont
			{Geremia}}, \bibinfo {author} {\bibfnamefont {A.~C.}\ \bibnamefont
			{Doherty}}, \ and\ \bibinfo {author} {\bibfnamefont {H.}~\bibnamefont
			{Mabuchi}},\ }\href {\doibase 10.1103/PhysRevA.67.022112} {\bibfield
		{journal} {\bibinfo  {journal} {Phys. Rev. A}\ }\textbf {\bibinfo {volume}
			{67}},\ \bibinfo {pages} {022112} (\bibinfo {year} {2003})}\BibitemShut
	{NoStop}%
	\bibitem [{\citenamefont {T{\'o}th}\ \emph {et~al.}(2010)\citenamefont
		{T{\'o}th}, \citenamefont {Wieczorek}, \citenamefont {Gross}, \citenamefont
		{Krischek}, \citenamefont {Schwemmer},\ and\ \citenamefont
		{Weinfurter}}]{toth2010}%
	\BibitemOpen
	\bibfield  {author} {\bibinfo {author} {\bibfnamefont {G.}~\bibnamefont
			{T{\'o}th}}, \bibinfo {author} {\bibfnamefont {W.}~\bibnamefont {Wieczorek}},
		\bibinfo {author} {\bibfnamefont {D.}~\bibnamefont {Gross}}, \bibinfo
		{author} {\bibfnamefont {R.}~\bibnamefont {Krischek}}, \bibinfo {author}
		{\bibfnamefont {C.}~\bibnamefont {Schwemmer}}, \ and\ \bibinfo {author}
		{\bibfnamefont {H.}~\bibnamefont {Weinfurter}},\ }\href {\doibase
		10.1103/PhysRevLett.105.250403} {\bibfield  {journal} {\bibinfo  {journal}
			{Phys. Rev. Lett.}\ }\textbf {\bibinfo {volume} {105}},\ \bibinfo {pages}
		{250403} (\bibinfo {year} {2010})}\BibitemShut {NoStop}%
	\bibitem [{\citenamefont {L{\"u}cke}\ \emph {et~al.}(2014)\citenamefont
		{L{\"u}cke}, \citenamefont {Peise}, \citenamefont {Vitagliano}, \citenamefont
		{Arlt}, \citenamefont {Santos}, \citenamefont {T{\'o}th},\ and\ \citenamefont
		{Klempt}}]{lucke2014}%
	\BibitemOpen
	\bibfield  {author} {\bibinfo {author} {\bibfnamefont {B.}~\bibnamefont
			{L{\"u}cke}}, \bibinfo {author} {\bibfnamefont {J.}~\bibnamefont {Peise}},
		\bibinfo {author} {\bibfnamefont {G.}~\bibnamefont {Vitagliano}}, \bibinfo
		{author} {\bibfnamefont {J.}~\bibnamefont {Arlt}}, \bibinfo {author}
		{\bibfnamefont {L.}~\bibnamefont {Santos}}, \bibinfo {author} {\bibfnamefont
			{G.}~\bibnamefont {T{\'o}th}}, \ and\ \bibinfo {author} {\bibfnamefont
			{C.}~\bibnamefont {Klempt}},\ }\href {\doibase
		10.1103/PhysRevLett.112.155304} {\bibfield  {journal} {\bibinfo  {journal}
			{Phys. Rev. Lett.}\ }\textbf {\bibinfo {volume} {112}},\ \bibinfo {pages}
		{155304} (\bibinfo {year} {2014})}\BibitemShut {NoStop}%
	\bibitem [{\citenamefont {D{\"u}r}\ \emph {et~al.}(2000)\citenamefont
		{D{\"u}r}, \citenamefont {Vidal},\ and\ \citenamefont {Cirac}}]{DVC00}%
	\BibitemOpen
	\bibfield  {author} {\bibinfo {author} {\bibfnamefont {W.}~\bibnamefont
			{D{\"u}r}}, \bibinfo {author} {\bibfnamefont {G.}~\bibnamefont {Vidal}}, \
		and\ \bibinfo {author} {\bibfnamefont {J.~I.}\ \bibnamefont {Cirac}},\ }\href
	{\doibase 10.1103/PhysRevA.62.062314} {\bibfield  {journal} {\bibinfo
			{journal} {Phys. Rev. A}\ }\textbf {\bibinfo {volume} {62}},\ \bibinfo
		{pages} {062314} (\bibinfo {year} {2000})}\BibitemShut {NoStop}%
	\bibitem [{\citenamefont {Sakurai}(1994)}]{sakurai1995modern}%
	\BibitemOpen
	\bibfield  {author} {\bibinfo {author} {\bibfnamefont {J.~J.}\ \bibnamefont
			{Sakurai}},\ }\href@noop {} {\emph {\bibinfo {title} {{Modern Quantum
					Mechanics}}}},\ \bibinfo {edition} {rev.}\ ed.\ (\bibinfo  {publisher}
	{Addison-Wesley, Reading},\ \bibinfo {year} {1994})\BibitemShut {NoStop}%
	\bibitem [{\citenamefont {Schwinger}(1965)}]{schwinger65}%
	\BibitemOpen
	\bibfield  {author} {\bibinfo {author} {\bibfnamefont {J.}~\bibnamefont
			{Schwinger}},\ }in\ \href@noop {} {\emph {\bibinfo {booktitle} {{Quantum
					Theory of Angular Momentum}}}},\ \bibinfo {editor} {edited by\ \bibinfo
		{editor} {\bibfnamefont {L.~C.}\ \bibnamefont {Biedenharn}}\ and\ \bibinfo
		{editor} {\bibfnamefont {H.}~\bibnamefont {Van~Dam}}}\ (\bibinfo  {publisher}
	{Academic Press, New York},\ \bibinfo {year} {1965})\ pp.\ \bibinfo {pages}
	{229--279}\BibitemShut {NoStop}%
	\bibitem [{\citenamefont {Brif}\ and\ \citenamefont {Mann}(1999)}]{Brif98}%
	\BibitemOpen
	\bibfield  {author} {\bibinfo {author} {\bibfnamefont {C.}~\bibnamefont
			{Brif}}\ and\ \bibinfo {author} {\bibfnamefont {A.}~\bibnamefont {Mann}},\
	}\href {\doibase 10.1103/PhysRevA.59.971} {\bibfield  {journal} {\bibinfo
			{journal} {Phys. Rev. A}\ }\textbf {\bibinfo {volume} {59}},\ \bibinfo
		{pages} {971} (\bibinfo {year} {1999})}\BibitemShut {NoStop}%
	\bibitem [{\citenamefont {Agarwal}(1981)}]{Agarwal81}%
	\BibitemOpen
	\bibfield  {author} {\bibinfo {author} {\bibfnamefont {G.~S.}\ \bibnamefont
			{Agarwal}},\ }\href {\doibase 10.1103/PhysRevA.24.2889} {\bibfield  {journal}
		{\bibinfo  {journal} {Phys. Rev. A}\ }\textbf {\bibinfo {volume} {24}},\
		\bibinfo {pages} {2889} (\bibinfo {year} {1981})}\BibitemShut {NoStop}%
	\bibitem [{\citenamefont {Dowling}\ \emph {et~al.}(1994)\citenamefont
		{Dowling}, \citenamefont {Agarwal},\ and\ \citenamefont
		{Schleich}}]{DowlingAgarwalSchleich}%
	\BibitemOpen
	\bibfield  {author} {\bibinfo {author} {\bibfnamefont {J.~P.}\ \bibnamefont
			{Dowling}}, \bibinfo {author} {\bibfnamefont {G.~S.}\ \bibnamefont
			{Agarwal}}, \ and\ \bibinfo {author} {\bibfnamefont {W.~P.}\ \bibnamefont
			{Schleich}},\ }\href {\doibase 10.1103/PhysRevA.49.4101} {\bibfield
		{journal} {\bibinfo  {journal} {Phys. Rev. A}\ }\textbf {\bibinfo {volume}
			{49}},\ \bibinfo {pages} {4101} (\bibinfo {year} {1994})}\BibitemShut
	{NoStop}%
	\bibitem [{\citenamefont {Heiss}\ and\ \citenamefont
		{Weigert}(2000)}]{heiss2000discrete}%
	\BibitemOpen
	\bibfield  {author} {\bibinfo {author} {\bibfnamefont {S.}~\bibnamefont
			{Heiss}}\ and\ \bibinfo {author} {\bibfnamefont {S.}~\bibnamefont
			{Weigert}},\ }\href {\doibase 10.1103/PhysRevA.63.012105} {\bibfield
		{journal} {\bibinfo  {journal} {Phys. Rev. A}\ }\textbf {\bibinfo {volume}
			{63}},\ \bibinfo {pages} {012105} (\bibinfo {year} {2000})}\BibitemShut
	{NoStop}%
	\bibitem [{\citenamefont {Klimov}\ and\ \citenamefont
		{de~Guise}(2010)}]{KdG10}%
	\BibitemOpen
	\bibfield  {author} {\bibinfo {author} {\bibfnamefont {A.~B.}\ \bibnamefont
			{Klimov}}\ and\ \bibinfo {author} {\bibfnamefont {H.}~\bibnamefont
			{de~Guise}},\ }\href {\doibase 10.1088/1751-8113/43/40/402001} {\bibfield
		{journal} {\bibinfo  {journal} {J. Phys. A}\ }\textbf {\bibinfo {volume}
			{43}},\ \bibinfo {pages} {402001} (\bibinfo {year} {2010})}\BibitemShut
	{NoStop}%
	\bibitem [{\citenamefont {Tilma}\ \emph {et~al.}(2016)\citenamefont {Tilma},
		\citenamefont {Everitt}, \citenamefont {Samson}, \citenamefont {Munro},\ and\
		\citenamefont {Nemoto}}]{tilma2016}%
	\BibitemOpen
	\bibfield  {author} {\bibinfo {author} {\bibfnamefont {T.}~\bibnamefont
			{Tilma}}, \bibinfo {author} {\bibfnamefont {M.~J.}\ \bibnamefont {Everitt}},
		\bibinfo {author} {\bibfnamefont {J.~H.}\ \bibnamefont {Samson}}, \bibinfo
		{author} {\bibfnamefont {W.~J.}\ \bibnamefont {Munro}}, \ and\ \bibinfo
		{author} {\bibfnamefont {K.}~\bibnamefont {Nemoto}},\ }\href {\doibase
		10.1103/PhysRevLett.117.180401} {\bibfield  {journal} {\bibinfo  {journal}
			{Phys. Rev. Lett.}\ }\textbf {\bibinfo {volume} {117}},\ \bibinfo {pages}
		{180401} (\bibinfo {year} {2016})}\BibitemShut {NoStop}%
	\bibitem [{\citenamefont {Rundle}\ \emph {et~al.}(2017)\citenamefont {Rundle},
		\citenamefont {Mills}, \citenamefont {Tilma}, \citenamefont {Samson},\ and\
		\citenamefont {Everitt}}]{rundle2017}%
	\BibitemOpen
	\bibfield  {author} {\bibinfo {author} {\bibfnamefont {R.~P.}\ \bibnamefont
			{Rundle}}, \bibinfo {author} {\bibfnamefont {P.~W.}\ \bibnamefont {Mills}},
		\bibinfo {author} {\bibfnamefont {T.}~\bibnamefont {Tilma}}, \bibinfo
		{author} {\bibfnamefont {J.~H.}\ \bibnamefont {Samson}}, \ and\ \bibinfo
		{author} {\bibfnamefont {M.~J.}\ \bibnamefont {Everitt}},\ }\href {\doibase
		10.1103/PhysRevA.96.022117} {\bibfield  {journal} {\bibinfo  {journal} {Phys.
				Rev. A}\ }\textbf {\bibinfo {volume} {96}},\ \bibinfo {pages} {022117}
		(\bibinfo {year} {2017})}\BibitemShut {NoStop}%
	\bibitem [{\citenamefont {Rundle}\ \emph {et~al.}(2019)\citenamefont {Rundle},
		\citenamefont {Tilma}, \citenamefont {Samson}, \citenamefont {Dwyer},
		\citenamefont {Bishop},\ and\ \citenamefont {Everitt}}]{RTD17}%
	\BibitemOpen
	\bibfield  {author} {\bibinfo {author} {\bibfnamefont {R.~P.}\ \bibnamefont
			{Rundle}}, \bibinfo {author} {\bibfnamefont {T.}~\bibnamefont {Tilma}},
		\bibinfo {author} {\bibfnamefont {J.~H.}\ \bibnamefont {Samson}}, \bibinfo
		{author} {\bibfnamefont {V.~M.}\ \bibnamefont {Dwyer}}, \bibinfo {author}
		{\bibfnamefont {R.~F.}\ \bibnamefont {Bishop}}, \ and\ \bibinfo {author}
		{\bibfnamefont {M.~J.}\ \bibnamefont {Everitt}},\ }\href {\doibase
		10.1103/PhysRevA.99.012115} {\ \textbf {\bibinfo {volume} {99}},\ \bibinfo
		{pages} {012115} (\bibinfo {year} {2019})}\BibitemShut {NoStop}%
	\bibitem [{\citenamefont {Gazeau}(2009)}]{Gazeau}%
	\BibitemOpen
	\bibfield  {author} {\bibinfo {author} {\bibfnamefont {J.-P.}\ \bibnamefont
			{Gazeau}},\ }\href@noop {} {\emph {\bibinfo {title} {{Coherent States in
					Quantum Physics}}}}\ (\bibinfo  {publisher} {Wiley-VCH, Weinheim},\ \bibinfo
	{year} {2009})\BibitemShut {NoStop}%
	\bibitem [{\citenamefont {Moya-Cessa}\ and\ \citenamefont
		{Knight}(1993)}]{moya1993}%
	\BibitemOpen
	\bibfield  {author} {\bibinfo {author} {\bibfnamefont {H.}~\bibnamefont
			{Moya-Cessa}}\ and\ \bibinfo {author} {\bibfnamefont {P.~L.}\ \bibnamefont
			{Knight}},\ }\href {\doibase 10.1103/PhysRevA.48.2479} {\bibfield  {journal}
		{\bibinfo  {journal} {Phys. Rev. A}\ }\textbf {\bibinfo {volume} {48}},\
		\bibinfo {pages} {2479} (\bibinfo {year} {1993})}\BibitemShut {NoStop}%
	\bibitem [{\citenamefont {Koczor}\ \emph {et~al.}(2018)\citenamefont {Koczor},
		\citenamefont {vom Ende}, \citenamefont {de~Gosson}, \citenamefont {Glaser},\
		and\ \citenamefont {Zeier}}]{koczor2018}%
	\BibitemOpen
	\bibfield  {author} {\bibinfo {author} {\bibfnamefont {B.}~\bibnamefont
			{Koczor}}, \bibinfo {author} {\bibfnamefont {F.}~\bibnamefont {vom Ende}},
		\bibinfo {author} {\bibfnamefont {M.~A.}\ \bibnamefont {de~Gosson}}, \bibinfo
		{author} {\bibfnamefont {S.~J.}\ \bibnamefont {Glaser}}, \ and\ \bibinfo
		{author} {\bibfnamefont {R.}~\bibnamefont {Zeier}},\ }\href@noop {} {\enquote
		{\bibinfo {title} {{Phase Spaces, Parity Operators, and the Born-Jordan
					Distribution}},}\ } (\bibinfo {year} {2018}),\ \bibinfo {note}
	{(\emph{Preprint} \href{https://arxiv.org/abs/1811.05872}{\tt
			arXiv:1811.05872})}\BibitemShut {NoStop}%
	\bibitem [{\citenamefont {Grossmann}(1976)}]{Grossmann1976}%
	\BibitemOpen
	\bibfield  {author} {\bibinfo {author} {\bibfnamefont {A.}~\bibnamefont
			{Grossmann}},\ }\href {\doibase 10.1007/BF01617867} {\bibfield  {journal}
		{\bibinfo  {journal} {Comm. Math. Phys.}\ }\textbf {\bibinfo {volume} {48}},\
		\bibinfo {pages} {191} (\bibinfo {year} {1976})}\BibitemShut {NoStop}%
	\bibitem [{\citenamefont {Messiah}(1962)}]{messiah1962}%
	\BibitemOpen
	\bibfield  {author} {\bibinfo {author} {\bibfnamefont {A.}~\bibnamefont
			{Messiah}},\ }\href@noop {} {\emph {\bibinfo {title} {{Quantum
					Mechanics~II}}}}\ (\bibinfo  {publisher} {North-Holland Publishing Company,
		Amsterdam},\ \bibinfo {year} {1962})\BibitemShut {NoStop}%
	\bibitem [{\citenamefont {Perelomov}(2012)}]{perelomov2012}%
	\BibitemOpen
	\bibfield  {author} {\bibinfo {author} {\bibfnamefont {A.}~\bibnamefont
			{Perelomov}},\ }\href@noop {} {\emph {\bibinfo {title} {{Generalized Coherent
					States and Their Applications}}}}\ (\bibinfo  {publisher} {Springer,
		Berlin},\ \bibinfo {year} {2012})\BibitemShut {NoStop}%
	\bibitem [{\citenamefont {Arecchi}\ \emph {et~al.}(1972)\citenamefont
		{Arecchi}, \citenamefont {Courtens}, \citenamefont {Gilmore},\ and\
		\citenamefont {Thomas}}]{arecchi1972atomic}%
	\BibitemOpen
	\bibfield  {author} {\bibinfo {author} {\bibfnamefont {F.}~\bibnamefont
			{Arecchi}}, \bibinfo {author} {\bibfnamefont {E.}~\bibnamefont {Courtens}},
		\bibinfo {author} {\bibfnamefont {R.}~\bibnamefont {Gilmore}}, \ and\
		\bibinfo {author} {\bibfnamefont {H.}~\bibnamefont {Thomas}},\ }\href
	{\doibase 10.1103/PhysRevA.6.2211} {\bibfield  {journal} {\bibinfo  {journal}
			{Phys. Rev. A}\ }\textbf {\bibinfo {volume} {6}},\ \bibinfo {pages} {2211}
		(\bibinfo {year} {1972})}\BibitemShut {NoStop}%
	\bibitem [{\citenamefont {Racah}(1942)}]{Racah42}%
	\BibitemOpen
	\bibfield  {author} {\bibinfo {author} {\bibfnamefont {G.}~\bibnamefont
			{Racah}},\ }\href {\doibase 10.1103/PhysRev.62.438} {\bibfield  {journal}
		{\bibinfo  {journal} {Phys. Rev.}\ }\textbf {\bibinfo {volume} {62}},\
		\bibinfo {pages} {438} (\bibinfo {year} {1942})}\BibitemShut {NoStop}%
	\bibitem [{\citenamefont {Koczor}(2019)}]{thesis}%
	\BibitemOpen
	\bibfield  {author} {\bibinfo {author} {\bibfnamefont {B.}~\bibnamefont
			{Koczor}},\ }\emph {\bibinfo {title} {{On phase-space representations of spin
				systems and their relations to infinite-dimensional quantum states}}},\ \href
	{http://mediatum.ub.tum.de?id=1463517} {\bibinfo {type} {Dissertation}},\
	\bibinfo  {school} {Technische Universität München}, \bibinfo {address}
	{Munich} (\bibinfo {year} {2019})\BibitemShut {NoStop}%
	\bibitem [{\citenamefont {Amiet}\ and\ \citenamefont
		{Weigert}(2000)}]{amiet2000contracting}%
	\BibitemOpen
	\bibfield  {author} {\bibinfo {author} {\bibfnamefont {J.-P.}\ \bibnamefont
			{Amiet}}\ and\ \bibinfo {author} {\bibfnamefont {S.}~\bibnamefont
			{Weigert}},\ }\href {\doibase 10.1103/PhysRevA.63.012102} {\bibfield
		{journal} {\bibinfo  {journal} {Phys. Rev. A}\ }\textbf {\bibinfo {volume}
			{63}},\ \bibinfo {pages} {012102} (\bibinfo {year} {2000})}\BibitemShut
	{NoStop}%
	\bibitem [{\citenamefont {Klimov}\ \emph
		{et~al.}(2017{\natexlab{b}})\citenamefont {Klimov}, \citenamefont {Romero},\
		and\ \citenamefont {de~Guise}}]{klimovgeneralized}%
	\BibitemOpen
	\bibfield  {author} {\bibinfo {author} {\bibfnamefont {A.~B.}\ \bibnamefont
			{Klimov}}, \bibinfo {author} {\bibfnamefont {J.~L.}\ \bibnamefont {Romero}},
		\ and\ \bibinfo {author} {\bibfnamefont {H.}~\bibnamefont {de~Guise}},\
	}\href {\doibase 10.1088/1751-8121/50/32/323001} {\bibfield  {journal}
		{\bibinfo  {journal} {J. Phys. A}\ }\textbf {\bibinfo {volume} {50}},\
		\bibinfo {pages} {1} (\bibinfo {year} {2017}{\natexlab{b}})}\BibitemShut
	{NoStop}%
	\bibitem [{\citenamefont {Stratonovich}(1956)}]{stratonovich1956}%
	\BibitemOpen
	\bibfield  {author} {\bibinfo {author} {\bibfnamefont {R.}~\bibnamefont
			{Stratonovich}},\ }\href@noop {} {\bibfield  {journal} {\bibinfo  {journal}
			{Sov. Phys. D}\ }\textbf {\bibinfo {volume} {1}},\ \bibinfo {pages} {414}
		(\bibinfo {year} {1956})}\BibitemShut {NoStop}%
	\bibitem [{\citenamefont {V{\'a}rilly}\ and\ \citenamefont
		{Garcia-Bond{\'i}a}(1989)}]{vgb89}%
	\BibitemOpen
	\bibfield  {author} {\bibinfo {author} {\bibfnamefont {J.~C.}\ \bibnamefont
			{V{\'a}rilly}}\ and\ \bibinfo {author} {\bibfnamefont {J.~M.}\ \bibnamefont
			{Garcia-Bond{\'i}a}},\ }\href {\doibase 10.1016/0003-4916(89)90262-5}
	{\bibfield  {journal} {\bibinfo  {journal} {Ann. Phys.}\ }\textbf {\bibinfo
			{volume} {190}},\ \bibinfo {pages} {107} (\bibinfo {year}
		{1989})}\BibitemShut {NoStop}%
	\bibitem [{\citenamefont {Koczor}\ \emph
		{et~al.}(2019{\natexlab{b}})\citenamefont {Koczor}, \citenamefont {Zeier},\
		and\ \citenamefont {Glaser}}]{koczor2016}%
	\BibitemOpen
	\bibfield  {author} {\bibinfo {author} {\bibfnamefont {B.}~\bibnamefont
			{Koczor}}, \bibinfo {author} {\bibfnamefont {R.}~\bibnamefont {Zeier}}, \
		and\ \bibinfo {author} {\bibfnamefont {S.~J.}\ \bibnamefont {Glaser}},\
	}\href {\doibase 10.1016/j.aop.2018.11.020} {\bibfield  {journal} {\bibinfo
			{journal} {Ann. Phys.}\ }\textbf {\bibinfo {volume} {408}},\ \bibinfo {pages}
		{1} (\bibinfo {year} {2019}{\natexlab{b}})}\BibitemShut {NoStop}%
	\bibitem [{\citenamefont {Tilma}\ and\ \citenamefont {Nemoto}(2012)}]{TN2012}%
	\BibitemOpen
	\bibfield  {author} {\bibinfo {author} {\bibfnamefont {T.}~\bibnamefont
			{Tilma}}\ and\ \bibinfo {author} {\bibfnamefont {K.}~\bibnamefont {Nemoto}},\
	}\href {\doibase 10.1088/1751-8113/45/1/015302} {\bibfield  {journal}
		{\bibinfo  {journal} {J. Phys. A}\ }\textbf {\bibinfo {volume} {45}},\
		\bibinfo {pages} {015302} (\bibinfo {year} {2012})}\BibitemShut {NoStop}%
	\bibitem [{\citenamefont {Garon}\ \emph {et~al.}(2015)\citenamefont {Garon},
		\citenamefont {Zeier},\ and\ \citenamefont {Glaser}}]{DROPS}%
	\BibitemOpen
	\bibfield  {author} {\bibinfo {author} {\bibfnamefont {A.}~\bibnamefont
			{Garon}}, \bibinfo {author} {\bibfnamefont {R.}~\bibnamefont {Zeier}}, \ and\
		\bibinfo {author} {\bibfnamefont {S.~J.}\ \bibnamefont {Glaser}},\ }\href
	{\doibase 10.1103/PhysRevA.91.042122} {\bibfield  {journal} {\bibinfo
			{journal} {Phys. Rev. A}\ }\textbf {\bibinfo {volume} {91}},\ \bibinfo
		{pages} {042122} (\bibinfo {year} {2015})}\BibitemShut {NoStop}%
	\bibitem [{\citenamefont {Brif}\ and\ \citenamefont {Mann}(1997)}]{Brif97}%
	\BibitemOpen
	\bibfield  {author} {\bibinfo {author} {\bibfnamefont {C.}~\bibnamefont
			{Brif}}\ and\ \bibinfo {author} {\bibfnamefont {A.}~\bibnamefont {Mann}},\
	}\href {\doibase 10.1088/0305-4470/31/1/002} {\bibfield  {journal} {\bibinfo
			{journal} {J. Phys. A}\ }\textbf {\bibinfo {volume} {31}},\ \bibinfo {pages}
		{L9} (\bibinfo {year} {1997})}\BibitemShut {NoStop}%
	\bibitem [{\citenamefont {Biedenharn}\ and\ \citenamefont
		{Louck}(1981)}]{BL81}%
	\BibitemOpen
	\bibfield  {author} {\bibinfo {author} {\bibfnamefont {L.~C.}\ \bibnamefont
			{Biedenharn}}\ and\ \bibinfo {author} {\bibfnamefont {J.~D.}\ \bibnamefont
			{Louck}},\ }\href@noop {} {\emph {\bibinfo {title} {{Angular Momentum in
					Quantum Physics}}}}\ (\bibinfo  {publisher} {Addison-Wesley, Reading, MA},\
	\bibinfo {year} {1981})\BibitemShut {NoStop}%
	\bibitem [{\citenamefont {Fano}(1953)}]{Fano53}%
	\BibitemOpen
	\bibfield  {author} {\bibinfo {author} {\bibfnamefont {U.}~\bibnamefont
			{Fano}},\ }\href {\doibase 10.1103/PhysRev.90.577} {\bibfield  {journal}
		{\bibinfo  {journal} {Phys. Rev.}\ }\textbf {\bibinfo {volume} {90}},\
		\bibinfo {pages} {577} (\bibinfo {year} {1953})}\BibitemShut {NoStop}%
	\bibitem [{\citenamefont {Tajima}(2015)}]{wigdmatrix1}%
	\BibitemOpen
	\bibfield  {author} {\bibinfo {author} {\bibfnamefont {N.}~\bibnamefont
			{Tajima}},\ }\href {\doibase 10.1103/PhysRevC.91.014320} {\bibfield
		{journal} {\bibinfo  {journal} {Phys. Rev. C}\ }\textbf {\bibinfo {volume}
			{91}},\ \bibinfo {pages} {014320} (\bibinfo {year} {2015})}\BibitemShut
	{NoStop}%
	\bibitem [{\citenamefont {Feng}\ \emph {et~al.}(2015)\citenamefont {Feng},
		\citenamefont {Wang}, \citenamefont {Yang},\ and\ \citenamefont
		{Jin}}]{wigdmatrix2}%
	\BibitemOpen
	\bibfield  {author} {\bibinfo {author} {\bibfnamefont {X.~M.}\ \bibnamefont
			{Feng}}, \bibinfo {author} {\bibfnamefont {P.}~\bibnamefont {Wang}}, \bibinfo
		{author} {\bibfnamefont {W.}~\bibnamefont {Yang}}, \ and\ \bibinfo {author}
		{\bibfnamefont {G.~R.}\ \bibnamefont {Jin}},\ }\href {\doibase
		10.1103/PhysRevE.92.043307} {\bibfield  {journal} {\bibinfo  {journal} {Phys.
				Rev. E}\ }\textbf {\bibinfo {volume} {92}},\ \bibinfo {pages} {043307}
		(\bibinfo {year} {2015})}\BibitemShut {NoStop}%
	\bibitem [{\citenamefont {Usha~Devi}\ \emph {et~al.}(2012)\citenamefont
		{Usha~Devi}, \citenamefont {Shudha},\ and\ \citenamefont
		{Rajagopal}}]{devi2012majorana}%
	\BibitemOpen
	\bibfield  {author} {\bibinfo {author} {\bibfnamefont {A.~R.}\ \bibnamefont
			{Usha~Devi}}, \bibinfo {author} {\bibnamefont {Shudha}}, \ and\ \bibinfo
		{author} {\bibfnamefont {A.~K.}\ \bibnamefont {Rajagopal}},\ }\href {\doibase
		10.1007/s11128-011-0280-8} {\bibfield  {journal} {\bibinfo  {journal}
			{Quantum Inf. Process.}\ }\textbf {\bibinfo {volume} {11}},\ \bibinfo {pages}
		{685} (\bibinfo {year} {2012})}\BibitemShut {NoStop}%
	\bibitem [{\citenamefont {Bj{\"o}rk}\ \emph {et~al.}(2015)\citenamefont
		{Bj{\"o}rk}, \citenamefont {Klimov}, \citenamefont {de~la Hoz}, \citenamefont
		{Grassl}, \citenamefont {Leuchs},\ and\ \citenamefont
		{S{\'a}nchez-Soto}}]{bjork2015}%
	\BibitemOpen
	\bibfield  {author} {\bibinfo {author} {\bibfnamefont {G.}~\bibnamefont
			{Bj{\"o}rk}}, \bibinfo {author} {\bibfnamefont {A.~B.}\ \bibnamefont
			{Klimov}}, \bibinfo {author} {\bibfnamefont {P.}~\bibnamefont {de~la Hoz}},
		\bibinfo {author} {\bibfnamefont {M.}~\bibnamefont {Grassl}}, \bibinfo
		{author} {\bibfnamefont {G.}~\bibnamefont {Leuchs}}, \ and\ \bibinfo {author}
		{\bibfnamefont {L.~L.}\ \bibnamefont {S{\'a}nchez-Soto}},\ }\href {\doibase
		10.1103/PhysRevA.92.031801} {\bibfield  {journal} {\bibinfo  {journal}
			{Physical Review A}\ }\textbf {\bibinfo {volume} {92}},\ \bibinfo {pages}
		{031801} (\bibinfo {year} {2015})}\BibitemShut {NoStop}%
	\bibitem [{\citenamefont {Giraud}\ \emph {et~al.}(2008)\citenamefont {Giraud},
		\citenamefont {Braun},\ and\ \citenamefont {Braun}}]{giraud2008}%
	\BibitemOpen
	\bibfield  {author} {\bibinfo {author} {\bibfnamefont {O.}~\bibnamefont
			{Giraud}}, \bibinfo {author} {\bibfnamefont {P.}~\bibnamefont {Braun}}, \
		and\ \bibinfo {author} {\bibfnamefont {D.}~\bibnamefont {Braun}},\ }\href
	{\doibase 10.1103/PhysRevA.78.042112} {\bibfield  {journal} {\bibinfo
			{journal} {Phys. Rev. A}\ }\textbf {\bibinfo {volume} {78}},\ \bibinfo
		{pages} {042112} (\bibinfo {year} {2008})}\BibitemShut {NoStop}%
	\bibitem [{\citenamefont {Groemer}(1996)}]{groemer1996}%
	\BibitemOpen
	\bibfield  {author} {\bibinfo {author} {\bibfnamefont {H.}~\bibnamefont
			{Groemer}},\ }\href@noop {} {\emph {\bibinfo {title} {{Geometric Applications
					of Fourier Series and Spherical Harmonics}}}}\ (\bibinfo  {publisher}
	{Cambridge University Press, Cambridge},\ \bibinfo {year} {1996})\BibitemShut
	{NoStop}%
	\bibitem [{\citenamefont {Kennedy}\ and\ \citenamefont
		{Sadeghi}(2013)}]{kennedy2013book}%
	\BibitemOpen
	\bibfield  {author} {\bibinfo {author} {\bibfnamefont {R.~A.}\ \bibnamefont
			{Kennedy}}\ and\ \bibinfo {author} {\bibfnamefont {P.}~\bibnamefont
			{Sadeghi}},\ }\href@noop {} {\emph {\bibinfo {title} {{Hilbert Space Methods
					in Signal Processing}}}}\ (\bibinfo  {publisher} {Cambridge University Press,
		Cambridge},\ \bibinfo {year} {2013})\BibitemShut {NoStop}%
	\bibitem [{\citenamefont {Bu{\v{z}}ek}\ \emph {et~al.}(1995)\citenamefont
		{Bu{\v{z}}ek}, \citenamefont {Keitel},\ and\ \citenamefont
		{Knight}}]{buvzek1995}%
	\BibitemOpen
	\bibfield  {author} {\bibinfo {author} {\bibfnamefont {V.}~\bibnamefont
			{Bu{\v{z}}ek}}, \bibinfo {author} {\bibfnamefont {C.~H.}\ \bibnamefont
			{Keitel}}, \ and\ \bibinfo {author} {\bibfnamefont {P.~L.}\ \bibnamefont
			{Knight}},\ }\href {\doibase 10.1103/PhysRevA.51.2594} {\bibfield  {journal}
		{\bibinfo  {journal} {Phys. Rev. A}\ }\textbf {\bibinfo {volume} {51}},\
		\bibinfo {pages} {2594} (\bibinfo {year} {1995})}\BibitemShut {NoStop}%
	\bibitem [{\citenamefont {Opatrn{\`y}}\ \emph {et~al.}(1995)\citenamefont
		{Opatrn{\`y}}, \citenamefont {Bu{\v{z}}ek}, \citenamefont {Bajer},\ and\
		\citenamefont {Drobn{\`y}}}]{opatrny1995}%
	\BibitemOpen
	\bibfield  {author} {\bibinfo {author} {\bibfnamefont {T.}~\bibnamefont
			{Opatrn{\`y}}}, \bibinfo {author} {\bibfnamefont {V.}~\bibnamefont
			{Bu{\v{z}}ek}}, \bibinfo {author} {\bibfnamefont {J.}~\bibnamefont {Bajer}},
		\ and\ \bibinfo {author} {\bibfnamefont {G.}~\bibnamefont {Drobn{\`y}}},\
	}\href {\doibase 10.1103/PhysRevA.52.2419} {\bibfield  {journal} {\bibinfo
			{journal} {Phys. Rev. A}\ }\textbf {\bibinfo {volume} {52}},\ \bibinfo
		{pages} {2419} (\bibinfo {year} {1995})}\BibitemShut {NoStop}%
	\bibitem [{\citenamefont {{Keih\"anen, E.}}\ and\ \citenamefont {{Reinecke,
				M.}}(2012)}]{beamdeconv}%
	\BibitemOpen
	\bibfield  {author} {\bibinfo {author} {\bibnamefont {{Keih\"anen, E.}}}\
		and\ \bibinfo {author} {\bibnamefont {{Reinecke, M.}}},\ }\href {\doibase
		10.1051/0004-6361/201220183} {\bibfield  {journal} {\bibinfo  {journal}
			{Astron. Astrophys.}\ }\textbf {\bibinfo {volume} {548}},\ \bibinfo {pages}
		{A110} (\bibinfo {year} {2012})}\BibitemShut {NoStop}%
	\bibitem [{\citenamefont {Wandelt}\ and\ \citenamefont
		{G\'orski}(2001)}]{fastconv}%
	\BibitemOpen
	\bibfield  {author} {\bibinfo {author} {\bibfnamefont {B.~D.}\ \bibnamefont
			{Wandelt}}\ and\ \bibinfo {author} {\bibfnamefont {K.~M.}\ \bibnamefont
			{G\'orski}},\ }\href {\doibase 10.1103/PhysRevD.63.123002} {\bibfield
		{journal} {\bibinfo  {journal} {Phys. Rev. D}\ }\textbf {\bibinfo {volume}
			{63}},\ \bibinfo {pages} {123002} (\bibinfo {year} {2001})}\BibitemShut
	{NoStop}%
	\bibitem [{Note1()}]{Note1}%
	\BibitemOpen
	\bibinfo {note} {The depicted random pure state vector is approximately given
		by $( 0.06 + i0.02,\protect \tmspace +\thinmuskip {.1667em} -0.21 -
		i0.19,\protect \tmspace +\thinmuskip {.1667em} 0.04 + i0.27,\protect \tmspace
		+\thinmuskip {.1667em} 0.15 - i0.11,\protect \tmspace +\thinmuskip {.1667em}
		0.28 - i0.28,\protect \tmspace +\thinmuskip {.1667em} -0.33 - i0.25,\protect
		\tmspace +\thinmuskip {.1667em} 0.04 - i0.44,\protect \tmspace +\thinmuskip
		{.1667em} -0.21 - i0.24,\protect \tmspace +\thinmuskip {.1667em} -0.43 +
		i0.00 )^T$.}\BibitemShut {Stop}%
	\bibitem [{\citenamefont {Ma}\ \emph {et~al.}(2011)\citenamefont {Ma},
		\citenamefont {Wang}, \citenamefont {Sun},\ and\ \citenamefont
		{Nori}}]{ma2011quantum}%
	\BibitemOpen
	\bibfield  {author} {\bibinfo {author} {\bibfnamefont {J.}~\bibnamefont
			{Ma}}, \bibinfo {author} {\bibfnamefont {X.}~\bibnamefont {Wang}}, \bibinfo
		{author} {\bibfnamefont {C.-P.}\ \bibnamefont {Sun}}, \ and\ \bibinfo
		{author} {\bibfnamefont {F.}~\bibnamefont {Nori}},\ }\href {\doibase
		10.1016/j.physrep.2011.08.003} {\bibfield  {journal} {\bibinfo  {journal}
			{Phys. Rep.}\ }\textbf {\bibinfo {volume} {509}},\ \bibinfo {pages} {89}
		(\bibinfo {year} {2011})}\BibitemShut {NoStop}%
	\bibitem [{\citenamefont {Deleglise}\ \emph {et~al.}(2008)\citenamefont
		{Deleglise}, \citenamefont {Dotsenko}, \citenamefont {Sayrin}, \citenamefont
		{Bernu}, \citenamefont {Brune}, \citenamefont {Raimond},\ and\ \citenamefont
		{Haroche}}]{deleglise2008}%
	\BibitemOpen
	\bibfield  {author} {\bibinfo {author} {\bibfnamefont {S.}~\bibnamefont
			{Deleglise}}, \bibinfo {author} {\bibfnamefont {I.}~\bibnamefont {Dotsenko}},
		\bibinfo {author} {\bibfnamefont {C.}~\bibnamefont {Sayrin}}, \bibinfo
		{author} {\bibfnamefont {J.}~\bibnamefont {Bernu}}, \bibinfo {author}
		{\bibfnamefont {M.}~\bibnamefont {Brune}}, \bibinfo {author} {\bibfnamefont
			{J.-M.}\ \bibnamefont {Raimond}}, \ and\ \bibinfo {author} {\bibfnamefont
			{S.}~\bibnamefont {Haroche}},\ }\href {\doibase 10.1038/nature07288}
	{\bibfield  {journal} {\bibinfo  {journal} {Nature}\ }\textbf {\bibinfo
			{volume} {455}},\ \bibinfo {pages} {510} (\bibinfo {year}
		{2008})}\BibitemShut {NoStop}%
	\bibitem [{\citenamefont {Lutterbach}\ and\ \citenamefont
		{Davidovich}(1997)}]{Lutterbach97}%
	\BibitemOpen
	\bibfield  {author} {\bibinfo {author} {\bibfnamefont {L.~G.}\ \bibnamefont
			{Lutterbach}}\ and\ \bibinfo {author} {\bibfnamefont {L.}~\bibnamefont
			{Davidovich}},\ }\href {\doibase 10.1103/PhysRevLett.78.2547} {\bibfield
		{journal} {\bibinfo  {journal} {Phys. Rev. Lett.}\ }\textbf {\bibinfo
			{volume} {78}},\ \bibinfo {pages} {2547} (\bibinfo {year}
		{1997})}\BibitemShut {NoStop}%
	\bibitem [{\citenamefont {Bertet}\ \emph {et~al.}(2002)\citenamefont {Bertet},
		\citenamefont {Auffeves}, \citenamefont {Maioli}, \citenamefont {Osnaghi},
		\citenamefont {Meunier}, \citenamefont {Brune}, \citenamefont {Raimond},\
		and\ \citenamefont {Haroche}}]{Bertet02}%
	\BibitemOpen
	\bibfield  {author} {\bibinfo {author} {\bibfnamefont {P.}~\bibnamefont
			{Bertet}}, \bibinfo {author} {\bibfnamefont {A.}~\bibnamefont {Auffeves}},
		\bibinfo {author} {\bibfnamefont {P.}~\bibnamefont {Maioli}}, \bibinfo
		{author} {\bibfnamefont {S.}~\bibnamefont {Osnaghi}}, \bibinfo {author}
		{\bibfnamefont {T.}~\bibnamefont {Meunier}}, \bibinfo {author} {\bibfnamefont
			{M.}~\bibnamefont {Brune}}, \bibinfo {author} {\bibfnamefont {J.~M.}\
			\bibnamefont {Raimond}}, \ and\ \bibinfo {author} {\bibfnamefont
			{S.}~\bibnamefont {Haroche}},\ }\href {\doibase
		10.1103/PhysRevLett.89.200402} {\bibfield  {journal} {\bibinfo  {journal}
			{Phys. Rev. Lett.}\ }\textbf {\bibinfo {volume} {89}},\ \bibinfo {pages}
		{200402} (\bibinfo {year} {2002})}\BibitemShut {NoStop}%
	\bibitem [{\citenamefont {Banaszek}\ \emph {et~al.}(1999)\citenamefont
		{Banaszek}, \citenamefont {Radzewicz}, \citenamefont {W\'odkiewicz},\ and\
		\citenamefont {Krasi\ifmmode~\acute{n}\else \'{n}\fi{}ski}}]{Banaszek99}%
	\BibitemOpen
	\bibfield  {author} {\bibinfo {author} {\bibfnamefont {K.}~\bibnamefont
			{Banaszek}}, \bibinfo {author} {\bibfnamefont {C.}~\bibnamefont {Radzewicz}},
		\bibinfo {author} {\bibfnamefont {K.}~\bibnamefont {W\'odkiewicz}}, \ and\
		\bibinfo {author} {\bibfnamefont {J.~S.}\ \bibnamefont
			{Krasi\ifmmode~\acute{n}\else \'{n}\fi{}ski}},\ }\href {\doibase
		10.1103/PhysRevA.60.674} {\bibfield  {journal} {\bibinfo  {journal} {Phys.
				Rev. A}\ }\textbf {\bibinfo {volume} {60}},\ \bibinfo {pages} {674} (\bibinfo
		{year} {1999})}\BibitemShut {NoStop}%
	\bibitem [{\citenamefont {Man'ko}\ and\ \citenamefont
		{Man'ko}(1997)}]{Manko97}%
	\BibitemOpen
	\bibfield  {author} {\bibinfo {author} {\bibfnamefont {V.~I.}\ \bibnamefont
			{Man'ko}}\ and\ \bibinfo {author} {\bibfnamefont {O.~V.}\ \bibnamefont
			{Man'ko}},\ }\href {\doibase 10.1134/1.558326} {\bibfield  {journal}
		{\bibinfo  {journal} {J. Exp. Theor. Phys.}\ }\textbf {\bibinfo {volume}
			{85}},\ \bibinfo {pages} {430} (\bibinfo {year} {1997})}\BibitemShut
	{NoStop}%
	\bibitem [{\citenamefont {{\.Z}yczkowski}\ \emph {et~al.}(2011)\citenamefont
		{{\.Z}yczkowski}, \citenamefont {Penson}, \citenamefont {Nechita},\ and\
		\citenamefont {Collins}}]{zyczkowski2011}%
	\BibitemOpen
	\bibfield  {author} {\bibinfo {author} {\bibfnamefont {K.}~\bibnamefont
			{{\.Z}yczkowski}}, \bibinfo {author} {\bibfnamefont {K.~A.}\ \bibnamefont
			{Penson}}, \bibinfo {author} {\bibfnamefont {I.}~\bibnamefont {Nechita}}, \
		and\ \bibinfo {author} {\bibfnamefont {B.}~\bibnamefont {Collins}},\ }\href
	{\doibase 10.1063/1.3595693} {\bibfield  {journal} {\bibinfo  {journal} {J.
				Math. Phys.}\ }\textbf {\bibinfo {volume} {52}},\ \bibinfo {pages} {062201}
		(\bibinfo {year} {2011})}\BibitemShut {NoStop}%
	\bibitem [{\citenamefont {Driscoll}\ and\ \citenamefont
		{Healy}(1994)}]{driscoll94}%
	\BibitemOpen
	\bibfield  {author} {\bibinfo {author} {\bibfnamefont {J.~R.}\ \bibnamefont
			{Driscoll}}\ and\ \bibinfo {author} {\bibfnamefont {D.~M.}\ \bibnamefont
			{Healy}},\ }\href {\doibase 10.1006/aama.1994.1008} {\bibfield  {journal}
		{\bibinfo  {journal} {Adv. Appl. Math.}\ }\textbf {\bibinfo {volume} {15}},\
		\bibinfo {pages} {202} (\bibinfo {year} {1994})}\BibitemShut {NoStop}%
	\bibitem [{\citenamefont {Lebedev}(1975)}]{leb75}%
	\BibitemOpen
	\bibfield  {author} {\bibinfo {author} {\bibfnamefont {V.~I.}\ \bibnamefont
			{Lebedev}},\ }\href {\doibase 10.1016/0041-5553(75)90133-0} {\bibfield
		{journal} {\bibinfo  {journal} {USSR Comput. Math. \& Math. Phys.}\ }\textbf
		{\bibinfo {volume} {15}},\ \bibinfo {pages} {44} (\bibinfo {year}
		{1975})}\BibitemShut {NoStop}%
	\bibitem [{\citenamefont {Lebedev}(1976)}]{leb76}%
	\BibitemOpen
	\bibfield  {author} {\bibinfo {author} {\bibfnamefont {V.~I.}\ \bibnamefont
			{Lebedev}},\ }\href {\doibase 10.1016/0041-5553(76)90100-2} {\bibfield
		{journal} {\bibinfo  {journal} {USSR Comput. Math. \& Math. Phys.}\ }\textbf
		{\bibinfo {volume} {16}},\ \bibinfo {pages} {10} (\bibinfo {year}
		{1976})}\BibitemShut {NoStop}%
	\bibitem [{\citenamefont {Lebedev}\ and\ \citenamefont {Laikov}(1999)}]{leb99}%
	\BibitemOpen
	\bibfield  {author} {\bibinfo {author} {\bibfnamefont {V.~I.}\ \bibnamefont
			{Lebedev}}\ and\ \bibinfo {author} {\bibfnamefont {D.~N.}\ \bibnamefont
			{Laikov}},\ }\href@noop {} {\bibfield  {journal} {\bibinfo  {journal}
			{Russian Acad. Sci. Dokl. Math.}\ }\textbf {\bibinfo {volume} {59}},\
		\bibinfo {pages} {477} (\bibinfo {year} {1999})}\BibitemShut {NoStop}%
	\bibitem [{\citenamefont {Leiner}\ \emph {et~al.}(2017)\citenamefont {Leiner},
		\citenamefont {Zeier},\ and\ \citenamefont {Glaser}}]{leiner17}%
	\BibitemOpen
	\bibfield  {author} {\bibinfo {author} {\bibfnamefont {D.}~\bibnamefont
			{Leiner}}, \bibinfo {author} {\bibfnamefont {R.}~\bibnamefont {Zeier}}, \
		and\ \bibinfo {author} {\bibfnamefont {S.~J.}\ \bibnamefont {Glaser}},\
	}\href {\doibase 10.1103/PhysRevA.96.063413} {\bibfield  {journal} {\bibinfo
			{journal} {Phys. Rev. A}\ }\textbf {\bibinfo {volume} {96}},\ \bibinfo
		{pages} {063413} (\bibinfo {year} {2017})}\BibitemShut {NoStop}%
	\bibitem [{\citenamefont {Leiner}\ and\ \citenamefont
		{Glaser}(2018)}]{Leiner18}%
	\BibitemOpen
	\bibfield  {author} {\bibinfo {author} {\bibfnamefont {D.}~\bibnamefont
			{Leiner}}\ and\ \bibinfo {author} {\bibfnamefont {S.~J.}\ \bibnamefont
			{Glaser}},\ }\href {\doibase 10.1103/PhysRevA.98.012112} {\bibfield
		{journal} {\bibinfo  {journal} {Phys. Rev. A}\ }\textbf {\bibinfo {volume}
			{98}},\ \bibinfo {pages} {012112} (\bibinfo {year} {2018})}\BibitemShut
	{NoStop}%
	\bibitem [{\citenamefont {Paris}\ and\ \citenamefont
		{\v{R}eh\'{a}\v{c}ek}(2004)}]{PR04}%
	\BibitemOpen
	\bibinfo {editor} {\bibfnamefont {M.}~\bibnamefont {Paris}}\ and\ \bibinfo
	{editor} {\bibfnamefont {J.}~\bibnamefont {\v{R}eh\'{a}\v{c}ek}},\ eds.,\
	\href@noop {} {\emph {\bibinfo {title} {{Quantum State Estimation}}}}\
	(\bibinfo  {publisher} {Springer, Berlin},\ \bibinfo {year}
	{2004})\BibitemShut {NoStop}%
	\bibitem [{\citenamefont {Amiet}\ and\ \citenamefont
		{Weigert}(1998)}]{amiet1998reconstructing}%
	\BibitemOpen
	\bibfield  {author} {\bibinfo {author} {\bibfnamefont {J.-P.}\ \bibnamefont
			{Amiet}}\ and\ \bibinfo {author} {\bibfnamefont {S.}~\bibnamefont
			{Weigert}},\ }\href {\doibase 10.1088/0305-4470/31/31/001} {\bibfield
		{journal} {\bibinfo  {journal} {J. Phys. A}\ }\textbf {\bibinfo {volume}
			{31}},\ \bibinfo {pages} {L543} (\bibinfo {year} {1998})}\BibitemShut
	{NoStop}%
	\bibitem [{\citenamefont {D’Ariano}\ \emph {et~al.}(2003)\citenamefont
		{D’Ariano}, \citenamefont {Maccone},\ and\ \citenamefont
		{Paini}}]{ariano2003spin}%
	\BibitemOpen
	\bibfield  {author} {\bibinfo {author} {\bibfnamefont {G.}~\bibnamefont
			{D’Ariano}}, \bibinfo {author} {\bibfnamefont {L.}~\bibnamefont {Maccone}},
		\ and\ \bibinfo {author} {\bibfnamefont {M.}~\bibnamefont {Paini}},\ }\href
	{\doibase 10.1088/1464-4266/5/1/311} {\bibfield  {journal} {\bibinfo
			{journal} {J. Opt. B}\ }\textbf {\bibinfo {volume} {5}},\ \bibinfo {pages}
		{77} (\bibinfo {year} {2003})}\BibitemShut {NoStop}%
	\bibitem [{\citenamefont {Schwemmer}\ \emph {et~al.}(2015)\citenamefont
		{Schwemmer}, \citenamefont {Knips}, \citenamefont {Richart}, \citenamefont
		{Weinfurter}, \citenamefont {Moroder}, \citenamefont {Kleinmann},\ and\
		\citenamefont {G{\"u}hne}}]{schwemmer15}%
	\BibitemOpen
	\bibfield  {author} {\bibinfo {author} {\bibfnamefont {C.}~\bibnamefont
			{Schwemmer}}, \bibinfo {author} {\bibfnamefont {L.}~\bibnamefont {Knips}},
		\bibinfo {author} {\bibfnamefont {D.}~\bibnamefont {Richart}}, \bibinfo
		{author} {\bibfnamefont {H.}~\bibnamefont {Weinfurter}}, \bibinfo {author}
		{\bibfnamefont {T.}~\bibnamefont {Moroder}}, \bibinfo {author} {\bibfnamefont
			{M.}~\bibnamefont {Kleinmann}}, \ and\ \bibinfo {author} {\bibfnamefont
			{O.}~\bibnamefont {G{\"u}hne}},\ }\href {\doibase
		10.1103/PhysRevLett.114.080403} {\bibfield  {journal} {\bibinfo  {journal}
			{Phys. Rev. Lett.}\ }\textbf {\bibinfo {volume} {114}},\ \bibinfo {pages}
		{080403} (\bibinfo {year} {2015})}\BibitemShut {NoStop}%
	\bibitem [{\citenamefont {Knips}\ \emph {et~al.}(2015)\citenamefont {Knips},
		\citenamefont {Schwemmer}, \citenamefont {Klein}, \citenamefont {Reuter},
		\citenamefont {T{\'o}th},\ and\ \citenamefont {Weinfurter}}]{knips15}%
	\BibitemOpen
	\bibfield  {author} {\bibinfo {author} {\bibfnamefont {L.}~\bibnamefont
			{Knips}}, \bibinfo {author} {\bibfnamefont {C.}~\bibnamefont {Schwemmer}},
		\bibinfo {author} {\bibfnamefont {N.}~\bibnamefont {Klein}}, \bibinfo
		{author} {\bibfnamefont {J.}~\bibnamefont {Reuter}}, \bibinfo {author}
		{\bibfnamefont {G.}~\bibnamefont {T{\'o}th}}, \ and\ \bibinfo {author}
		{\bibfnamefont {H.}~\bibnamefont {Weinfurter}},\ }\href@noop {} {\enquote
		{\bibinfo {title} {{How long does it take to obtain a physical density
					matrix?}}}\ } (\bibinfo {year} {2015}),\ \Eprint
	{http://arxiv.org/abs/1512.06866v1} {arXiv:1512.06866v1} \BibitemShut
	{NoStop}%
	\bibitem [{\citenamefont {Faist}\ and\ \citenamefont {Renner}(2016)}]{faist16}%
	\BibitemOpen
	\bibfield  {author} {\bibinfo {author} {\bibfnamefont {P.}~\bibnamefont
			{Faist}}\ and\ \bibinfo {author} {\bibfnamefont {R.}~\bibnamefont {Renner}},\
	}\href {\doibase 10.1103/PhysRevLett.117.010404} {\bibfield  {journal}
		{\bibinfo  {journal} {Phys. Rev. Lett.}\ }\textbf {\bibinfo {volume} {117}},\
		\bibinfo {pages} {010404} (\bibinfo {year} {2016})}\BibitemShut {NoStop}%
	\bibitem [{\citenamefont {Silva}\ \emph {et~al.}(2017)\citenamefont {Silva},
		\citenamefont {Glancy},\ and\ \citenamefont {Vasconcelos}}]{silva17}%
	\BibitemOpen
	\bibfield  {author} {\bibinfo {author} {\bibfnamefont {G.~B.}\ \bibnamefont
			{Silva}}, \bibinfo {author} {\bibfnamefont {S.}~\bibnamefont {Glancy}}, \
		and\ \bibinfo {author} {\bibfnamefont {H.~M.}\ \bibnamefont {Vasconcelos}},\
	}\href {\doibase 10.1103/PhysRevA.95.022107} {\bibfield  {journal} {\bibinfo
			{journal} {Phys. Rev. A}\ }\textbf {\bibinfo {volume} {95}},\ \bibinfo
		{pages} {022107} (\bibinfo {year} {2017})}\BibitemShut {NoStop}%
	\bibitem [{\citenamefont {Steffens}\ \emph {et~al.}(2017)\citenamefont
		{Steffens}, \citenamefont {Riofr{\'i}}, \citenamefont {McCutcheon},
		\citenamefont {Roth}, \citenamefont {Bell}, \citenamefont {McMillan},
		\citenamefont {Tame}, \citenamefont {Rarity},\ and\ \citenamefont
		{Eisert}}]{steffens17}%
	\BibitemOpen
	\bibfield  {author} {\bibinfo {author} {\bibfnamefont {A.}~\bibnamefont
			{Steffens}}, \bibinfo {author} {\bibfnamefont {C.~A.}\ \bibnamefont
			{Riofr{\'i}}}, \bibinfo {author} {\bibfnamefont {W.}~\bibnamefont
			{McCutcheon}}, \bibinfo {author} {\bibfnamefont {I.}~\bibnamefont {Roth}},
		\bibinfo {author} {\bibfnamefont {B.~A.}\ \bibnamefont {Bell}}, \bibinfo
		{author} {\bibfnamefont {A.}~\bibnamefont {McMillan}}, \bibinfo {author}
		{\bibfnamefont {M.~S.}\ \bibnamefont {Tame}}, \bibinfo {author}
		{\bibfnamefont {J.~G.}\ \bibnamefont {Rarity}}, \ and\ \bibinfo {author}
		{\bibfnamefont {J.}~\bibnamefont {Eisert}},\ }\href {\doibase
		10.1088/2058-9565/aa6ae2} {\bibfield  {journal} {\bibinfo  {journal} {Quantum
				Sci. Technol.}\ }\textbf {\bibinfo {volume} {2}},\ \bibinfo {pages} {025005}
		(\bibinfo {year} {2017})}\BibitemShut {NoStop}%
	\bibitem [{\citenamefont {Riofr{\'i}}\ \emph {et~al.}(2017)\citenamefont
		{Riofr{\'i}}, \citenamefont {Gross}, \citenamefont {Flammia}, \citenamefont
		{Monz}, \citenamefont {Nigg}, \citenamefont {Blatt},\ and\ \citenamefont
		{Eisert}}]{riofrio17}%
	\BibitemOpen
	\bibfield  {author} {\bibinfo {author} {\bibfnamefont {C.~A.}\ \bibnamefont
			{Riofr{\'i}}}, \bibinfo {author} {\bibfnamefont {D.}~\bibnamefont {Gross}},
		\bibinfo {author} {\bibfnamefont {S.~T.}\ \bibnamefont {Flammia}}, \bibinfo
		{author} {\bibfnamefont {T.}~\bibnamefont {Monz}}, \bibinfo {author}
		{\bibfnamefont {D.}~\bibnamefont {Nigg}}, \bibinfo {author} {\bibfnamefont
			{R.}~\bibnamefont {Blatt}}, \ and\ \bibinfo {author} {\bibfnamefont
			{J.}~\bibnamefont {Eisert}},\ }\href {\doibase 10.1038/ncomms15305}
	{\bibfield  {journal} {\bibinfo  {journal} {Nat. Commun.}\ }\textbf {\bibinfo
			{volume} {8}},\ \bibinfo {pages} {15305} (\bibinfo {year}
		{2017})}\BibitemShut {NoStop}%
	\bibitem [{\citenamefont {Suess}\ \emph {et~al.}(2017)\citenamefont {Suess},
		\citenamefont {Rudnicki}, \citenamefont {Gross},\ and\ \citenamefont
		{Maciel}}]{suess16}%
	\BibitemOpen
	\bibfield  {author} {\bibinfo {author} {\bibfnamefont {D.}~\bibnamefont
			{Suess}}, \bibinfo {author} {\bibfnamefont {{\L}.}~\bibnamefont {Rudnicki}},
		\bibinfo {author} {\bibfnamefont {D.}~\bibnamefont {Gross}}, \ and\ \bibinfo
		{author} {\bibfnamefont {T.~O.}\ \bibnamefont {Maciel}},\ }\href {\doibase
		10.1088/1367-2630/aa7ce9} {\bibfield  {journal} {\bibinfo  {journal} {New J.
				Phys.}\ }\textbf {\bibinfo {volume} {19}},\ \bibinfo {pages} {093013}
		(\bibinfo {year} {2017})}\BibitemShut {NoStop}%
	\bibitem [{\citenamefont {Schervish}(1995)}]{schervish}%
	\BibitemOpen
	\bibfield  {author} {\bibinfo {author} {\bibfnamefont {M.~J.}\ \bibnamefont
			{Schervish}},\ }\href@noop {} {\emph {\bibinfo {title} {{Theory of
					Statistics}}}}\ (\bibinfo  {publisher} {Springer, New York},\ \bibinfo {year}
	{1995})\BibitemShut {NoStop}%
\end{thebibliography}

%

\end{document}